\pdfoutput=1
\documentclass [a4paper, 12pt]{article}
\usepackage[inner=2cm,outer=2cm,tmargin=2.5cm,bmargin=2.5cm]{geometry}
\usepackage{mathtools}
\usepackage{amssymb}
\usepackage[utf8]{inputenc}
\usepackage[T1]{fontenc}
\usepackage[english]{babel}
\usepackage{parskip}
\usepackage{tensor}
\usepackage{hyperref}
\usepackage{amsmath}
\usepackage{afterpage}
\usepackage{ae,aecompl}
\usepackage{graphicx}
\usepackage[capposition=bottom]{floatrow}
\usepackage{empheq}
\usepackage{amsthm}
\usepackage{lipsum} 
\usepackage{cite}
\usepackage{cases}
\usepackage{hyperref}
\usepackage{pgfplots}
\allowdisplaybreaks

\hypersetup{
	colorlinks,
	linkcolor={red!80!black},
	citecolor={blue!80!black},
	urlcolor={blue!80!black},
	linktocpage=page
}

\usetikzlibrary{decorations.pathmorphing,calc}


\providecommand{\abs}[1]{\lvert#1\rvert} 

\newcommand{\sd}{_\sigma}
\newcommand{\su}{^\sigma}
\newcommand{\rd}{_\rho}
\newcommand{\rup}{^\rho}

\newcommand{\mup}{^{\mu}}
\newcommand{\mud}{_{\mu}}
\newcommand{\nup}{^{\nu}}
\newcommand{\nud}{_{\nu}}
\newcommand{\au}{^{\alpha}}
\newcommand{\ad}{_{\alpha}}

\newcommand{\munuu}{^{\mu\nu}}
\newcommand{\munud}{_{\mu\nu}}
\newcommand{\abu}{^{\alpha\beta}}
\newcommand{\abd}{_{\alpha\beta}}
\newcommand{\nn}{\nonumber}

\newcommand{\pt}{\partial}

\newcommand{\na}{\nabla}

\newcommand{\ABU}{^{AB}}
\newcommand{\ABD}{_{AB}}

\newcommand{\mone}{^{-1}}
\newcommand{\mtwo}{^{-2}}
\newcommand{\eps}{\epsilon}

\newcommand{\nr}[1]{(\ref{#1})}  
\newcommand{\rmi}[1]{{\mbox{\scriptsize #1}}}  

\newcommand{\order}[1]{\mathcal{O}\left(#1\right)} 
\newcommand{\pminus}[1]{^{-#1}}

\numberwithin{equation}{section}

\newcommand{\be}{\begin{equation}}
\newcommand{\ee}{\end{equation}}
\newcommand{\bea}{\begin{eqnarray}}
\newcommand{\eea}{\end{eqnarray}}

\begin{document}
	
\begin{flushright}
HIP-2023-3/TH\\
\end{flushright}
\vspace{1cm}
	
\begin{center}
		
\centerline{\Large {\bf Gravitational wave memory in conformally flat spacetimes}}
		
\vspace{8mm}

\renewcommand\thefootnote{\mbox{$\fnsymbol{footnote}$}}
Niko Jokela,${}^{1,2}$\footnote{niko.jokela@helsinki.fi}
K. Kajantie,${}^2$\footnote{keijo.kajantie@helsinki.fi} and
Miika Sarkkinen${}^{1}$\footnote{miika.sarkkinen@helsinki.fi}
\vspace{4mm}

${}^1${\small \sl Department of Physics} and ${}^2${\small \sl Helsinki Institute of Physics} \\
{\small \sl P.O.Box 64} \\
{\small \sl FIN-00014 University of Helsinki, Finland} 
		
\vspace{0.8cm}
\end{center}
	
\vspace{0.8cm}
	
\setcounter{footnote}{0}
\renewcommand\thefootnote{\mbox{\arabic{footnote}}}

\begin{abstract}
\noindent 
We study the gravitational wave memory effect in spacetimes related to flat space by a conformal transformation. The discussion is general but the gravitational wave length scale is assumed to be small compared with the background curvature radius. The general formulas are applied to Friedmann--Robertson--Walker metrics of all spatial curvatures. The effect of new terms stemming from spatial curvature is potentially detectable in future gravitational wave measurements.
\end{abstract}
	
\newpage

\section{Introduction}

Gravitational wave memory effect is one of the intriguing predictions of general relativity (GR) \cite{Zeldovich:1974gvh,Braginsky:1986ia,Christodoulou:1991cr,Wiseman:1991ss,Thorne:1992sdb}. It is a persistent displacement between observers in geodesic motion, a permanent imprint of gravitational waves in spacetime geometry. While its existence in GR has been known for decades, it is yet to be confirmed experimentally. Although the currently operating instruments are not highly adapted for measuring the memory effect, its observation in the near future seems quite possible \cite{Lasky:2016knh,Grant:2022bla}.
	
In \cite{Bieri:2013ada}, Bieri and Garfinkle showed that the memory effect consists of two pieces: null memory sourced by stress-energy that gets to null infinity, on the one hand, and ordinary memory due to sources that travel to the future timelike infinity, on the other. 
Their treatment of the memory effect was based on the analysis of the Weyl tensor perturbations, which can be used to determine the shift in the detectors via geodesic deviation equation. An advantage of their approach was that in linear perturbation theory, perturbations of the Weyl tensor are gauge-invariant, due to the fact that in the Minkowski spacetime curvature along with its Weyl part vanishes. In this fashion, by working immediately with physical quantities, one need not worry about potential spurious gauge modes in one's results. In \cite{Bieri:2015jwa}, the method was applied in the slightly more nontrivial geometry of de Sitter spacetime.
	
To take full advantage of the method of \cite{Bieri:2013ada}, our purpose here is to generalize the analysis carried out therein and study the memory effect in a more general conformally flat background geometry where the Weyl tensor vanishes. This class of geometries is of special interest since many physically important spacetimes, including Minkowski spacetime and various cosmological spacetimes, are conformally flat. For a background with identically vanishing Weyl curvature, the Weyl tensor is exhausted by its perturbations, which are manifestly gauge-invariant. The Weyl tensor is the part of Riemann curvature that is not algebraically dependent on stress-energy and hence involves the true gravitational degrees of freedom; it thus encodes a freely propagating gravitational field. It is therefore very convenient to use the Weyl tensor to study gravitational waves and their memory effect in a conformally flat background. For an early study of cosmological perturbations in terms of the Weyl tensor, see \cite{Hawking:1966qi}.
	
We also apply our results to one example of a conformally flat spacetime: a Friedmann-Robertson-Walker (FRW) spacetime with arbitrary spatial curvature. The FRW metric describes a globally hyperbolic manifold with homogeneous and isotropic three-dimensional time slices of constant spatial curvature. This is a cosmologically interesting metric, the spatially flat special case being a very good approximation to our universe as far as latest measurements go \cite{Planck:2018vyg}. However, despite the stringent observational constraints on spatial curvature, it might be of importance to find new, independent observables that can be used to set further constraints on it. Since the memory effect will be a new GW observable in the future, it is therefore interesting to study the memory in an FRW metric with arbitrary spatial curvature. Also, this might be interesting from a purely theoretical or geometrical perspective as well even if not providing further experimental avenue for determining the spatial geometry of our universe.
	
The paper is structured as follows. We first review some basic properties of the Weyl tensor and derive the equations of motion and constraints satisfied by the curvature perturbations. Then we make an asymptotic ansatz for the perturbations, which furnishes us with a pair of equations that determine the memory effect. In the second part of the paper, we use our results to study the memory effect in a spatially curved FRW spacetime. We close with a discussion on these results. Two appendices are provided to lay out some supplementary calculations.

{\bf Conventions.} We use the geometric units $c = G = 1$ throughout the paper. For the Minkowski metric $\eta\munud$, we use the mostly plus $(-,+,+,+,\ldots,+)$ sign convention. Greek letters stand for spacetime indices, and Latin letters are used for spatial indices. In a spherical coordinate system, capital Latin letters \(A,B,C,\ldots\) denote angular indices. We denote symmetrization of tensor indices by $(\mu_1\cdots\mu_n)$ and antisymmetrization correspondingly $[\mu_1\cdots\mu_n]$ so that, \emph{e.g.}, $T\munud = T_{(\mu\nu)} + T_{[\mu\nu]}$.
We call the time coordinate of a conformally flat coordinate system `conformal time', whereas the time coordinate $\tau \equiv \int dt/a(t)$ in the FRW spacetime we call `parametric time.'

\section{General setup}\label{weylsec}
The Weyl tensor plays a pivotal role in our study of the memory effect, as alluded to in the introduction, we shall start this section by laying down some basic facts and identities satisfied by the tensor. We then gear our focus on the associated memory effect.

\subsection{Conformal flatness and the Weyl tensor}
	
Consider a conformally flat spacetime $(M, g)$ with dimension $n \geq 4$.
Conformal flatness is characterized by the condition that the metric is locally conformally equivalent to the Minkowski metric. More precisely, at each point on the manifold, we can find a coordinate neighborhood in which the metric reads
\begin{equation}\label{conf}
		g\munud(x) = \Omega^2(x) \, \eta\munud
\end{equation}
for some non-zero function $\Omega$, which we call a conformal factor. 
	
The Weyl tensor $C_{\mu\nu\sigma\rho}$ is the trace-free part of the Riemann tensor that appears in the Ricci decomposition of the latter: 
\begin{equation}\label{riccidec}
    R_{\mu\nu\rho\sigma} = C_{\mu\nu\rho\sigma} + \frac{2}{n-2}(g_{\mu[\rho}R_{\sigma]\nu} - g_{\nu[\rho}R_{\sigma]\mu})- \frac{2}{(n-1)(n-2)}g_{\mu[\rho}g_{\sigma]\nu}R \ .
\end{equation}
It has the same algebraic symmetries as the Riemann tensor:
\bea
 C_{\mu\nu\rho\sigma} & = & C_{[\mu\nu][\rho\sigma]} = C_{[\rho\sigma][\mu\nu]} \\
 C_{\mu[\nu\rho\sigma]} & = & 0 \ .
\eea
Due to the conformal invariance of the Weyl tensor, it vanishes identically in a conformally flat spacetime.
	
Given a unit timelike vector field $u\mup$, the Weyl tensor can be decomposed into the electric and magnetic parts:
\begin{equation}
		\tensor{C}{_\mu_\nu^\rho^\sigma} = 4 \delta_{[\mu}^{[\rho} \mathcal{E}_{\nu]}^{\sigma]} - 8 u_{[\mu}\tensor{\mathcal{E}}{_{\nu]}^{[\rho}}u^{\sigma]} + 2 \epsilon_{\mu\nu\alpha\beta}u^\alpha \mathcal{B}^{\beta[\rho}u^{\sigma]} + 2 \epsilon^{\rho\sigma\alpha\beta}u_\alpha \mathcal{B}_{\beta[\mu}u_{\nu]} \ ,
\end{equation}
where the electric part is
\begin{equation}\label{Eweyl}
	\mathcal{E}\munud = C_{\mu\rho\nu\sigma}u\rup u^\sigma
	\end{equation}
and the magnetic part is
\be\label{Bweyl}
	\mathcal{B}\munud = \star C_{\mu\rho\nu\sigma}u\rup u^\sigma = \frac{1}{2}\tensor{\epsilon}{_\mu_\rho^\alpha^\beta} C_{\alpha\beta\nu\sigma}u\rup u^\sigma \ .
\ee
Here $\star$ denotes the dual of the Weyl tensor, and $\epsilon$ is the spacetime volume form.
	
The Bianchi identity for the Weyl tensor reads
\be
	\na^\rho C_{\rho\sigma\mu\nu} = 2\, \frac{n-3}{n-2}\left[
	\na_{[\mu} R_{\nu]\sigma}+\frac{1}{2(n-1)}g_{\sigma[\mu}\na_{\nu]}R
	\right].
\ee
Using the Schouten tensor
\be
		S\munud = \frac{1}{n-2}\left[
		R\munud - \frac{R}{2(n-1)}g\munud
		\right]
\ee
the Bianchi identity can be written more compactly as
\be
	\na^\rho C_{\rho\sigma\mu\nu} = 2(n-3) \na_{[\mu} S_{\nu]\sigma} \ .
\ee
Einstein equation with the cosmological constant
\begin{equation}\label{einstein}
	R\munud -\frac{1}{2}R g\munud +\Lambda g\munud = 8 \pi T\munud
\end{equation}
can be used to relate the divergence of the Weyl tensor to stress-energy involved in spacetime:
\begin{equation}\label{weylEOM}
	\na^\rho C_{\rho\sigma\mu\nu} = 8 \pi \na_{[\mu} X_{\nu]\sigma} \ ,  
\end{equation}
where we have defined
\begin{equation}\label{matter}
	X\munud \equiv  2 \frac{n-3}{n-2} \left[
	T\munud - \frac{n-2}{2(n-1)}Tg\munud 
	\right], \quad T = g\munuu T\munud \ .
\end{equation}
The Weyl equation of motion \nr{weylEOM} is a GR analogue to Maxwell's equation of motion for the electromagnetic field
\be
 \na\mud F\munuu = J\nup \ ,
\ee  
where $F\munuu$ is the electromagnetic field strength and $J\nup$ is the $n$-current density. Acting now with $\na^\sigma$ on \nr{weylEOM}, we get on the LHS
\begin{equation}
	\na_\rho \na\sd \tensor{C}{^\rho^\sigma_\mu_\nu} = C_{\rho\sigma\alpha[\mu} \tensor{C}{^\rho^\sigma^\alpha_{\nu]}} \ .
\end{equation}
To derive this relation one has to use first the commutator rule for covariant derivatives, apply the Ricci decomposition \nr{riccidec}, and then use symmetry properties of the Weyl and Ricci tensors. But from the algebraic Bianchi identity for Weyl it follows that the tensor on the RHS is identically zero. Therefore, we get
\begin{equation}\label{divdivweyl}
	\na_\rho \na\sd \tensor{C}{^\rho^\sigma_\mu_\nu} = 0 \ .
\end{equation}
This is a gravitational analogue of the conservation law in electromagnetism:
\begin{equation}
	\na\nud\na\mud  F\munuu = \na\nud J\nup = 0 \ .
\end{equation}
To complete the analogue, we should now verify that also the RHS of \nr{weylEOM} vanishes under the second derivative. By direct computation: 
\begin{equation}\label{ddstress}
	\na\su \na_{[\mu} T_{\nu]\sigma} = \na_{[\mu} \na\su T_{\nu]\sigma} + \tensor{R}{^\sigma_{[\mu\nu]}^\rho}T_{\rho\sigma} + \tensor{R}{_{[\mu}^\sigma}T_{\nu]\sigma} = \tensor{R}{_{[\mu}^\sigma}T_{\nu]\sigma} = 8\pi T_{\sigma[\mu} \tensor{T}{_{\nu]}^\sigma} = 0 \ .
\end{equation}
In the first equality we again applied the commutator rule, and in the second one we used stress-energy conservation\footnote{Since the Weyl tensor vanishes in a conformally flat spacetime, the stress tensor satisfies the relation $\na_{[\mu} T_{\nu]}^\rho = \frac{n-2}{2(n-1)} \delta^\rho_{[\nu}\na_{\mu]} T$.
For $n=4$ this includes the energy-momentum conservation law but is overall a stronger constraint on the stress tensor. For $n>4$ this relation does not in general imply energy-momentum conservation, but instead imposing energy-momentum conservation sets the further constraint that the gradient of the trace of the stress tensor be vanishing, {\emph{i.e.}}, the trace must be a constant. Via Einstein's equation, the same holds for the Ricci scalar.} and symmetries of the Riemann tensor to get rid of the first two terms. Then, we used Einstein's equation \nr{einstein} where all terms proportional to the metric tensor give zero contribution due to the antisymmetrization. The final equality simply follows from symmetry of the stress tensor. Quite naturally, there also exists an analogy between gravitational and electromagnetic (EM) memory effects \cite{Bieri:2013hqa} (see \cite{Winicour:2014ska,Yoshida:2017fao,Mao:2017wvx,Hamada:2017atr,Sarkkinen:2018qwj,Campoleoni:2019ptc,Jokela:2019apz,Enriquez-Rojo:2022ntu,Taghiloo:2022adt,Bhatkar:2022qhz} for more discussion on EM memory effect).

\subsection{Weyl tensor perturbations}
Now we wish to perturb away from conformal flatness and to this end we use the notation $\bar f$ for background quantities and $\delta f$ for deviations thereof. From now on we will only consider dimension $n=4$.

Since the background Weyl tensor vanishes $\bar C_{\rho\sigma\mu\nu}=0$, the perturbation of the Weyl tensor is equal to the Weyl tensor of the perturbed spacetime itself. The Weyl tensor perturbation is hence gauge invariant and, along with the energy-momentum perturbation, satisfies the equation of motion (\ref{weylEOM}). Perturbing Eq. \nr{weylEOM} yields
\begin{equation}\label{weylpert}
	\bar{\na}\rup \delta C_{\rho\sigma\mu\nu} = 8\pi \left(
	\pt_{[\mu} \delta X_{\nu]\sigma} - \bar{\Gamma}^\rho_{\sigma[\mu} \delta X_{\nu]\rho} - \delta\Gamma^\rho_{\sigma[\mu}  \bar{X}_{\nu]\rho}
	\right) = 8\pi \left( \bar{\na}_{[\mu} \delta X_{\nu]\sigma} -  \delta\Gamma^\rho_{\sigma[\mu}  \bar{X}_{\nu]\rho}\right) \ ,
\end{equation}
where $\delta X$ is the perturbation of the source tensor defined in Eq. \nr{matter}. Along with this equation, we should also perturb the conservation law:
\begin{equation}\label{consv}
	\delta\left(\na\mup T\munud\right) = \bar{\na}\mup \delta T\munud + \delta g^{\rho\mu} \bar{\na}\rd \bar{T}\munud - \bar{g}^{\rho\mu}\delta \Gamma^\lambda_{\rho\mu} \bar{T}_{\lambda\nu} - \bar{g}^{\rho\mu}\delta \Gamma^\lambda_{\rho\nu} \bar{T}_{\mu\lambda} = 0 \ .
\end{equation}
These two equations form a coupled system for curvature and stress energy perturbations. In the presence of the coupling terms between background stress-energy and perturbations in the geometry in \nr{weylpert} and \nr{consv}, solving this system is in general difficult. However, things simplify a lot when we make an assumption that we are going to need also later: we assume that the perturbation length scale is small compared to the background curvature radius. This is just the standard separation of scales usually employed in the theory of gravitational radiation \cite{Maggiore:2007ulw}.
See Eq. \nr{separation} for a more detailed explanation.

Also note that the background matter tensor satisfies
\begin{equation}
	\pt_{[\mu} \bar{X}_{\nu]\sigma} = \bar{\Gamma}^\rho_{\sigma[\mu}  \bar{X}_{\nu]\rho} \ ,
\end{equation}
and also, using Einstein's equation,
\begin{equation}
\bar{X}\munud = \frac{1}{8\pi}\left(\bar{R}\munud - \frac{1}{3}\bar{g}\munud (\bar{R}+\Lambda)\right) \ .
\end{equation}
Due to Einstein's equation, the coupling term in \nr{weylpert} is $\sim (\pt \Omega)^2, \pt^2\Omega$, so assuming slowly varying background curvature, the coupling term is negligible compared to the covariant derivative of the matter perturbation, which is only proportional to $\pt\Omega$. The same holds for the coupling terms involving background stress--energy in \nr{consv}. Therefore, the system we are effectively solving is
\bea
  \bar{\na}\rup \delta C_{\rho\sigma\mu\nu} & = &  8\pi \bar{\na}_{[\mu} \delta X_{\nu]\sigma} \label{eomsimple} \\
  \bar{\na}\mup \delta T\munud & = & 0 \ .
\eea
Thus, we can separately solve for stress--energy perturbations propagating freely in the background spacetime, and then determine the curvature perturbations induced thereby. From now on, to avoid extra clutter we simply denote $C_{\rho\sigma\mu\nu} \equiv \delta C_{\rho\sigma\mu\nu}$, $T\munud \equiv \delta T\munud, X\munud \equiv \delta X\munud$.
	
Background Christoffel symbols in a conformal Cartesian frame read
\begin{equation}\label{cartchris}
	\bar{\Gamma}^\rho\munud = \delta^\rho\mud \na\nud \log \Omega + \delta^\rho\nud \na\mud \log \Omega - \eta\munud \eta^{\rho\lambda}\na_\lambda \log \Omega \ .
\end{equation}
In Cartesian coordinates, Eq. (\ref{eomsimple}) then becomes
\begin{equation}\label{weylCart}
	\pt^\rho \left(\Omega^{-1} C_{\rho\sigma\mu\nu} \right) = 8\pi \left[
		\Omega\, \pt_{[\mu} X_{\nu]\sigma} + X_{\sigma[\mu} \pt_{\nu]} \Omega + \eta_{\sigma[\mu} X_{\nu]\rho} \pt^\rho \Omega
		\right] \ ,
\end{equation}
where indices are raised with the Minkowski metric. It is then convenient to decompose the Weyl tensor into pieces that are proportional to its electric and magnetic parts defined in \nr{Eweyl} and \nr{Bweyl}:
\bea
	E_{ab} & = & \Omega^{-1} C_{a \eta b\eta} \\
	B_{ab} & = & \frac{1}{2}\Omega^{-1}\tensor{\epsilon}{^e^f_a} C_{efb\eta} \ ,
\eea
and the stress-energy tensor as follows:
\begin{equation}
	\mu = T_{\eta \eta}\ , \quad q_a = T_{\eta a}\ , \quad U_{ab} = T_{ab} \ .
\end{equation}
Here $\eta$ is the conformal time coordinate, Latin letters $a, b, c,\ldots$ denote the spatial components, and $\epsilon_{abc} = \epsilon_{\eta abc}$ with $\epsilon_{\mu\nu\sigma\rho}$ being the Levi-Civita tensor of Minkowski spacetime. From Eq. (\ref{weylCart}), we get the following two constraints:
\bea
	\pt^b E_{ab} & = & 4\pi \Omega \left[
	\frac{1}{3}\pt_a\left(
	2 \mu +  \tensor{U}{^c_c}
	\right) - \pt_\eta q_a - \frac{1}{3}\frac{\pt_a \Omega}{\Omega}
	\left(
	\mu + 2\tensor{U}{^c_c}
	\right) + \frac{\pt^b \Omega}{\Omega}U_{ab}
	\right] \label{Econstraint}
	\\
	\pt^b B_{ab} & = & 4 \pi \Omega\, 
	\tensor{\epsilon}{^e^f_a}
	\left[
    \pt_e q_f + q_e \, \pt_f \log\Omega
	\right] \label{Bconstraint}
\eea
and two equations of motion:
\begin{align}
	&\pt_\eta E_{ab} - \frac{1}{2}\tensor{\epsilon}{_a^c^d}\pt_c B_{db} -\frac{1}{2}\tensor{\epsilon}{_b^c^d}\pt_c B_{da} \nn \\
	= &4\pi\Omega \left[
	\pt_{(a} q_{b)} - \frac{1}{3}\delta_{ab} \pt_c q^c -\pt_\eta \left(
	U_{ab} - \frac{1}{3}\delta_{ab}\tensor{U}{^c_c}
	\right)
	\right]
	+ 4\pi \pt_\eta \Omega 
		\left[
		U_{ab} - \frac{1}{3}\delta_{ab}\tensor{U}{^c_c}
		\right] \nn \\
		&+ \frac{4\pi}{3}\delta_{ab} q^c\pt_c \Omega - 4\pi q_{(a} \pt_{b)}\Omega \\
		&\pt_\eta B_{ab} + \frac{1}{2}\tensor{\epsilon}{_a^c^d}\pt_c E_{db} + \frac{1}{2}\tensor{\epsilon}{_b^c^d}\pt_c E_{d a} \nn \\
		= &2 \pi \Omega \left[
		\tensor{\epsilon}{_a^c^d}\pt_c U_{db} + \tensor{\epsilon}{_b^c^d}\pt_c U_{da} + \tensor{\epsilon}{_a^c^d}U_{bc} \pt_d \log \Omega +
		\tensor{\epsilon}{_b^c^d}U_{ac} \pt_d \log \Omega 
		\right] \ .
\end{align}
Then we decompose the electric and magnetic parts of the Weyl tensor and the energy-momentum tensor into scalar, vector, and tensor parts in the spherical coordinates:
\begin{align}
	&E_{rr}, \quad B_{rr}, \quad q_r,\quad U_{rr}, \quad N = \tensor{U}{_c^c} \\
	&X_A = E_{rA}, \quad Y_A = B_{rA},\quad q_A, \quad V_A = U_{rA} \\
	&\tilde{E}_{AB} = E_{AB} - \frac{1}{2}h_{AB}\tensor{E}{_C^C}, \quad \tilde{B}_{AB} = B\ABD - \frac{1}{2}h\ABD \tensor{B}{_C^C}, \quad W\ABD = U\ABD -\frac{1}{2}h\ABD \tensor{U}{_C^C} \ ,
\end{align}
where the capital Latin letters are raised and lowered with the unit two-sphere metric $h\ABD$. The constraint equations now become:
\begin{align}
	&\pt_r E_{rr} + 3 r\mone E_{rr} + r\mtwo D^A X_A \nn \\
	= &4\pi \Omega 
		\left[
		\frac{1}{3}\pt_r \left(
		2\mu + N
		\right)
		-\pt_\eta q_r -\frac{1}{3} \frac{\pt_r \Omega}{\Omega}
		\left(
		\mu + 2 N
		\right)
		+ \frac{\pt_r \Omega}{\Omega} U_{rr} + r\mtwo V_A \frac{D^A \Omega}{\Omega}
		\right] \\
		&\pt_r B_{rr} + 3 r\mone B_{rr} + r\mtwo D^A Y_A \nn \\
		= & 4\pi \Omega r\mtwo \epsilon^{AB} 
		\left[
		D_A q_B + q_A D_B \log \Omega
		\right] \\
		&\pt_r X_A + 2 r\mone X_A - \frac{1}{2} D_A E_{rr} + r\mtwo D^B \tilde{E}_{BA} \nn \\
		= &4\pi \Omega 
		\left[
		\frac{1}{3}D_A \left(
		2\mu + N
		\right)
		-\pt_\eta q_A -\frac{1}{3} \frac{D_A \Omega}{\Omega}
		\left(
		\mu + 2 N
		\right)
		+ \frac{\pt_r \Omega}{\Omega} V_A + r\mtwo W_{AB} \frac{D^B \Omega}{\Omega} + \frac{1}{2}(N - U_{rr}) \frac{D_A \Omega}{\Omega}
		\right], \\
		&\pt_r Y_A + 2 r\mone Y_A - \frac{1}{2} D_A B_{rr} + r\mtwo D^B \tilde{B}_{BA} \nn \\
		= &4\pi \Omega \tensor{\epsilon}{_A^B}
		\left[
		D_B q_r - \pt_r q_B + q_B \pt_r \log \Omega - q_r D_B\log \Omega
		\right] \ .
\end{align}
Here $\epsilon\ABD$ is the volume form of the unit two-sphere. The equations of motion read as
\begin{align}
		&\pt_\eta E_{rr} - r\mtwo \epsilon\ABU D_A Y_B \nn \\
		= & 4\pi \Omega \left[
		\pt_r q_r - \pt_\eta U_{rr} + \frac{1}{3}\pt_\eta (N-\mu)
		\right]
		+ 4\pi \pt_\eta \Omega \left[
		U_{rr} - \frac{1}{3}(2 N +\mu)
		\right] + 4\pi r\mtwo q^A D_A \Omega \\
		&\pt_\eta B_{rr} + r\mtwo \epsilon \ABU D_A X_B \nn \\
		= & 4 \pi \Omega r\mtwo \epsilon\ABU \left[
		D_A V_B + V_A D_B \log \Omega 
		\right] \\
		&\pt_\eta X_A - \frac{1}{2}r\mtwo \epsilon^{BC} D_B \tilde{B}_{CA} - \frac{1}{4}\tensor{\epsilon}{_A^B}\left(
		3 D_B B_{rr} - 2 \pt_r Y_B
		\right) \nn \\
		= &2\pi\Omega \left[
		D_A q_r + \pt_r q_A
		\right] - 4\pi \Omega \left[
		r\mone q_A +\pt_\eta V_A
		\right] + 4 \pi \pt_\eta \Omega V_A - 2\pi \left[
		q_r D_A \Omega + q_A \pt_r \Omega
		\right] \\
		&\pt_\eta Y_A + \frac{1}{2}r\mtwo \epsilon^{BC} D_B \tilde{E}_{CA} +\frac{1}{4}\tensor{\epsilon}{_A^B}
		\left(
		3 D_B E_{rr} - 2 \pt_r X_B
		\right) \nn \\
		= & 2\pi \Omega \left[
		r\mtwo \epsilon^{BC} D_B W_{CA} + \frac{1}{2}\tensor{\epsilon}{_A^B} \left(
		D_B (3 U_{rr} - N) - \pt_r V_B
		\right) \right. \nn \\
		&\quad \quad \left. + r\mtwo \epsilon^{BC} W_{AB} D_C \log \Omega - \frac{1}{2}\tensor{\epsilon}{_A^B}( (3U_{rr} - N) D_B \log \Omega - V_B \pt_r \log \Omega )
		\right] \\
		&\pt_\eta \tilde{E}\ABD - \frac{1}{2}\tensor{\epsilon}{_A^C} \left(
		D_C Y_B - \pt_r \tilde{B}_{CB} + r\mone \tilde{B}_{CB}
		\right) \nn \\
		& - \frac{1}{2}\tensor{\epsilon}{_B^C} \left(
		D_C Y_A - \pt_r \tilde{B}_{CA} + r\mone \tilde{B}_{CA}
		\right) - \frac{1}{2}h\ABD \epsilon^{CD} D_C Y_D \nn \\
		=& 4 \pi \Omega \left[
		D_{(A} q_{B)}- \frac{1}{2}h\ABD D_C q^C x - \pt_\eta W\ABD
		\right] + 4\pi \pt_\eta \Omega W\ABD - 4\pi q_{(A} D_{B)}\Omega + 2\pi h\ABD q^C D_C \Omega \\
		&\pt_\eta \tilde{B}\ABD + \frac{1}{2}\tensor{\epsilon}{_A^C}\left(
		D_C X_B - \pt_r \tilde{E}_{CB} + r\mone \tilde{E}_{CB}
		\right) \nn \\
		& + \frac{1}{2}\tensor{\epsilon}{_B^C}\left(
		D_C X_A - \pt_r \tilde{E}_{CA} + r\mone \tilde{E}_{CA}
		\right) + \frac{1}{2}h\ABD \epsilon^{CD} D_C X_D \nn \\
		= & 2\pi \Omega \tensor{\epsilon}{_A^C} \left(
		D_C V_B + r\mone W_{CB} - \pt_r W_{CB}
		\right)
		+ 2\pi \Omega \tensor{\epsilon}{_B^C}\left(
		D_C V_A + r\mone W_{CA} - \pt_r W_{CA}
		\right) \nn \\
		&+ 2\pi \Omega \tensor{\epsilon}{_A^C}\left(
		W_{BC}\pt_r \log \Omega - V_B D_C \log \Omega
		\right)
		+ 2\pi \Omega \tensor{\epsilon}{_B^C}\left(
		W_{AC}\pt_r \log \Omega - V_A D_C \log \Omega
		\right) \nn \\
		&+ 2\pi \Omega h_{AB}\epsilon^{CD}\left(
		D_C V_D + V_C D_D \log \Omega
		\right) \ .
\end{align}

\subsection{Large $r$ asymptotics}

We shall now derive the asymptotic equations of motion and constraints. The calculation is similar to the one done in \cite{Bieri:2013ada} but as we will soon see, some new terms are going to emerge, which is why it is important we go through the derivation in some detail.
	
Consider stress-energy perturbation of the form
\begin{equation}\label{nullergy}
	T\munud = A\, \pt\mud u \,\pt\nud u \ , \quad u = \eta - r \ ,
\end{equation}
living in this background. This describes null energy traveling towards the future null infinity. Then the energy-momentum conservation law entails that
\begin{equation}
	0 = A \, \na\mup \na\mud u  + \na\mup A \na\mud u \ ,
\end{equation}
and solving this yields
\begin{equation}\label{Asol}
	A = \frac{L}{\Omega^2 r^2}
\end{equation}
for some $L = L(u,x^B)$; physically, this quantity is radiated power per unit solid angle. 
	
Suppose then that the Weyl tensor components fall off as
\bea
	\tilde{E}_{AB} & = & r \, e_{AB} + \ldots \label{asympt1}\\
	\tilde{B}_{AB} & = & r \, b_{AB} + \ldots \\
	X_A & = & x_A r^{-1} + \ldots \\
	Y_A & = & y_A r^{-1} + \ldots \\
	E_{rr} & = & P r^{-3} + \ldots \\
	B_{rr} & = & Q r^{-3} + \ldots \ . \label{asympt2}
\eea
We give an argument in support of these assumptions in Appendix \ref{app:weyl}.
Before and after the GW burst, $P$ is the only non-vanishing coefficient in the expansions \cite{Bieri:2013ada,Bieri:2015jwa}.
Then we plug these into the constraint equations, which become, to leading order,
\bea
		-\dot{P} + D^A x_A & = & -8 \pi \Omega^{-2} L (\Omega + r \pt_\eta \Omega + r \pt_r \Omega) \\
		-\dot{Q} + D^A y_A & = & 0 \\
		-\dot{x}_A + D^B e_{AB} & = & 0 \\
		-\dot{y}_A + D^B b_{BA} & = & 0 \ . \label{ybrelation}
\eea
Similarly, the equations of motion become
\bea
		\dot{P} - \epsilon^{AB} D_A y_B & = & 8 \pi \Omega^{-2} L\left( \Omega + r \pt_r \Omega + r\pt_\eta \Omega \right) \\
		\dot{Q} + \epsilon^{AB} D_A x_B & = & 0 \\
		\dot{x}_A - \frac{1}{2} \epsilon^{BC} D_B b_{CA} - \frac{1}{2}\tensor{\epsilon}{_A^B} \dot{y}_B & = & 0 \\
		\dot{y}_A + \frac{1}{2}\epsilon^{BC} D_B e_{CA} + \frac{1}{2}\epsilon_A^{\,\,\,\, B} \dot{x}_B & = & 0 \label{yexrelation}\\
		\dot{e}_{AB} - \tensor{\epsilon}{_A^C} \dot{b}_{CB} & = & 0 \label{ebrelation} \\
		\dot{b}_{AB} + \tensor{\epsilon}{_A^C} \dot{e}_{CB} & = & 0 \ .
\eea
Integrating Eq. \nr{ebrelation} over time and using the fact that before the burst $e\ABD, b\ABD \rightarrow 0$, we get $b\ABD = - \tensor{\epsilon}{_A^C} e_{CB}$. Then Eqs. \nr{ybrelation} and \nr{yexrelation} imply that $y_A = - \tensor{\epsilon}{_A^B} x_B$ since also $x_A, y_A \rightarrow 0$ in the past. We can then eliminate $y_A$ and $b_{AB}$ from the above equations; the remaining independent equations become
\bea
		D^A x_A & = & \dot{P} - 8 \pi L \Omega^{-2} \left( \Omega + r \pt_r \Omega + r \pt_\eta \Omega \right) \label{divx}\\
		D^B e_{AB} & = & \dot{x}_A \\
		\dot{Q} & = & - \epsilon^{AB} D_A x_B \\
		\epsilon^{BC}D_B e_{CA} & = & \tensor{\epsilon}{_A^B} \dot{x}_B \ .
\eea
We now define the following quantities:
\bea
		F & \equiv & \int_{u_i}^{u_f} L \,du\\
		v_{AB} & \equiv & \int_{u_i}^{u} e_{AB} du \\
		m_{AB} & \equiv & \int_{u_i}^{u_f} v_{AB} du \label{memorydef}\\
		z_A & \equiv & \int_{u_i}^{u_f} x_A du \ ,
\eea
where $F$ is the total energy radiated per unit solid angle, $v\ABD$ is called the velocity tensor, and $m\ABD$ is called the memory tensor. Note that all of these quantities are defined in the reference frame of a conformal observer.
Then we can write our equations as
\bea
		D^A z_A & = & \Delta P - 8 \pi F \Omega^{-2} \left( \Omega + r \pt_r \Omega + r \pt_\eta \Omega \right) \label{gauss}\\
		D^B v_{AB} & = & x_A \\
		\epsilon^{AB} D_A z_B & = & 0 \label{curlfree}\\
		\epsilon^{BC} D_B v_{CA} & = & \tensor{\epsilon}{_A^B} x_B \ .
\eea
In (\ref{gauss}) we assumed that $\Omega$ remains approximately constant over the burst of radiation. Eq. (\ref{curlfree}) implies that $z_A$ is irrotational: $z_A = D_A \phi$ for some function $\phi$ living on ${\mathbb{S}}^2$. Our final set of equations becomes
\bea
		D^A D_A \phi & = & \Delta P - 8 \pi F \Omega^{-2} \left( \Omega + r \pt_r \Omega + r \pt_\eta \Omega \right) \label{final1}\\
		D^B m_{AB} & = & D_A \phi \ . \label{final2}
\eea
This is a pair of equations for fields defined on the two-sphere; taking a divergence of the latter equation yields a relation between the memory tensor $m_{AB}$ and the source. We derive the solution in Appendix~\ref{app:memsol}; here we just cite the result:
\begin{equation}\label{memorysol1}
		m\ABD = \sum_{l\geq 2} m_{lm} \left(D_A D_B Y_{lm} - \frac{1}{2}h\ABD D_C D^C Y_{lm}\right) \ ,
\end{equation}
where the modes $m_{lm}$ are determined from the source by
\begin{equation}\label{memorysol2}
		m_{lm} = \frac{2}{(l-1)l(l+1)(l+2)}\int d\Omega \,Y_{lm}^* \left(\Delta P - 8\pi F\Omega^{-2} \left( \Omega + r \pt_r \Omega + r \pt_\eta \Omega \right) \right) \ .
\end{equation}
This is our first main result.
Here both ordinary and null contributions to the memory are present.
Once the memory tensor is known, one may use it in the geodesic deviation equation to determine the memory effect. Notice that the potential modifications to the memory effect due to a conformally flat background are encoded in the last two terms inside parentheses in \nr{final1}. The term $r \pt_\eta \Omega$ was already present in \cite{Bieri:2015jwa} where $\Omega$ was simply the cosmological scale factor $a(\eta)$. In our generalized treatment, a novel term $r \pt_r \Omega$ emerges, which leads to some new potentially observable modifications in the memory effect.

\subsection{Memory from geodesic deviation}

The metric becomes flat in the local inertial coordinates of an observer but curvature of spacetime cannot be removed by a change of coordinates. Effect of curvature on relative acceleration between two nearby geodesics is dictated by the geodesic deviation equation
\begin{equation}\label{deviation}
		u^\rho \na_\rho (u^\sigma \na_\sigma s\mup) =  \tensor{R}{^\mu_\nu_\rho_\sigma}u^\nu u^\rho s^\sigma \ ,
\end{equation}
where $u\mup$ is the unit tangent vector to one of the geodesics, and $s\mup$ is the separation vector pointing to the other geodesic. To get the memory effect experienced in a local observer frame, we go to the Fermi Normal Coordinates (FNC) centered around the worldline of the observer \cite{Poisson:2009pwt,Chu:2019ssw}. FNC are constructed by choosing a set of orthonormal basis vectors $\{e\mup_t,e\mup_a\}$ that are parallel transported along the observer worldline, and such that $e\mup_t = u\mup$. Here $\mu$ is the spacetime index associated with a basis vector and $a$ enumerates spacelike basis vectors. The proper time $t$ of the observer then gives the time coordinate of FNC, which is extended away from a point $p$ on the observer worldline by following spacelike geodesics that emanate in the perpendicular direction from $p$. A point $q$ lying on a spacelike geodesic at proper distance $\lambda$ from $p$ is then labeled by the coordinates $(t, \Omega^a \lambda)$ where $\Omega^a$ are the components of the unit tangent vector to the spacelike geodesic in the basis $\{e\mup_a\}$, evaluated at $p$. The coordinates remain well-defined as long as different spacelike geodesics emanating from the observer worldline do not cross each other elsewhere in the coordinate neighborhood.
	
The geodesic deviation equation in FNC reads
\begin{equation}
	\frac{d^2 s^a}{dt^2} = \tensor{R}{^a_t_t_b} s^b \ ,
\end{equation}
where
\begin{equation}\label{projection}
		\tensor{R}{^a_t_t_b} \equiv e^a\mud \tensor{R}{^\mu_\nu_\rho_\sigma} e\nup_t e^\rho_t e^\sigma_b
\end{equation}
is the projection of the Riemann tensor into the local observer frame, $e^a\mud$ being an element in the dual basis. In FNC the tensorial deviation equation hence becomes a system of ordinary differential equations for scalars living on the observer worldline. To obtain the memory effect, we integrate twice along the worldline:
\begin{equation}
		\Delta s^a = \int_{t_i}^{t_f}dt \int_{t_i}^{t}dt' \tensor{R}{^a_t_t_b} s^b \approx s^b \int_{t_i}^{t_f}dt \int_{t_i}^{t}dt' \tensor{R}{^a_t_t_b} \ .
\end{equation}
Using the Ricci decomposition \nr{riccidec}, contributions from different algebraic parts of the Riemann tensor become explicit:
\bea
    \Delta s^a & \approx & s^b \int_{t_i}^{t_f}dt \int_{t_i}^{t}dt' \left(
    \tensor{C}{^a_t_t_b} - \frac{1}{2} \delta^a_b R_{tt} - g_{t[t} \tensor{R}{_b_]^a} + \frac{1}{6} \delta^a_b g_{tt} R 
    \right) \nn \\
    & = & s^b \int_{t_i}^{t_f}dt \int_{t_i}^{t}dt' \left(
    \tensor{C}{^a_t_t_b} - \frac{1}{2} \delta^a_b R_{tt} + \frac{1}{2} \tensor{R}{_b^a} - \frac{1}{6} \delta^a_b  R 
    \right) \ .
\eea
Above we used the fact that in FNC the metric becomes diagonal on the worldline, and $g_{tt} = -1$.
Notice that the Ricci tensor perturbations fall off as fast as the stress-energy perturbations at large distances. Assuming that the background curvature radius is large, we may approximate that for a burst of short duration the Ricci part of the background Riemann tensor induces only a negligible tidal force in the small neighborhood around the observer worldline, so that the dominant effect comes from the ripples carried by the Weyl tensor. The GW memory effect is thus given by 
\begin{equation}\label{memeffect}
    \Delta s^a \approx s^b \int_{t_i}^{t_f}dt \int_{t_i}^{t}dt' \,
    \tensor{C}{^a_t_t_b} \ .
\end{equation}
Hence, given solutions for the Weyl tensor components in some chosen coordinate system, we then have to express the projection $\tensor{C}{^a_t_t_b} = e^a\mud \tensor{C}{^\mu_\nu_\rho_\sigma} e\nup_t e^\rho_t e^\sigma_b$ in terms of these solutions. Let us note that this holds for freely falling observers.

\section{Application: spatially curved FRW spacetime}
	
In this section, we will apply our general formulas derived in the previous section to the FRW spacetime with spatial curvature that can take either positive, negative, or zero value. To implement our results, we first need to find a conformally flat form of the curved FRW metric. After that, we will express the memory effect in the local frame of a comoving observer.
Here a curved FRW universe refers to a universe whose metric in comoving coordinates has constant spatial curvature. In other words, hypersurfaces parametrized by proper time of the cosmic fluid have constant Gaussian curvature $K$. We will take this local representation of the FRW metric as our starting point. 

The canonical foliation of the FRW spacetime, though natural in a sense, is still an arbitrary choice; one could reslice the manifold to obtain a spatial geometry one prefers. For instance, de Sitter spacetime can be foliated by hypersurfaces of zero, positive, or negative curvature. To apply the general formulas for the memory effect, the metric in the new foliation would then have to be transformed to the conformally flat form. One could then parametrize the effect measured by an observer in terms of the new foliation but the result would be the same as what we will get with the comoving foliation.

\subsection{Spatially curved FRW metric in conformally flat coordinates}
	
\begin{figure}[t!]
		\begin{center}
			\includegraphics[width=0.49\textwidth]{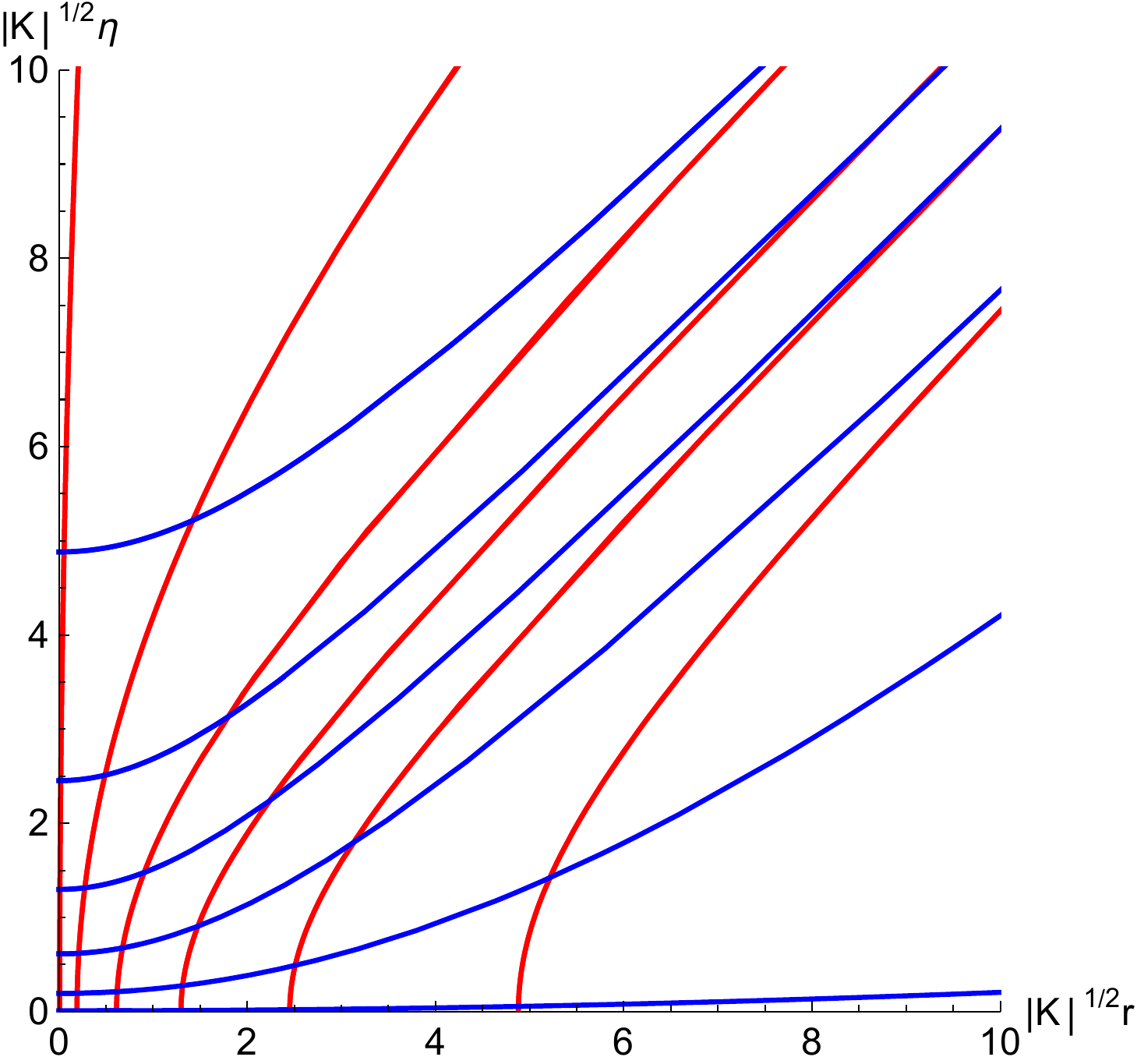}
			\includegraphics[width=0.49\textwidth]{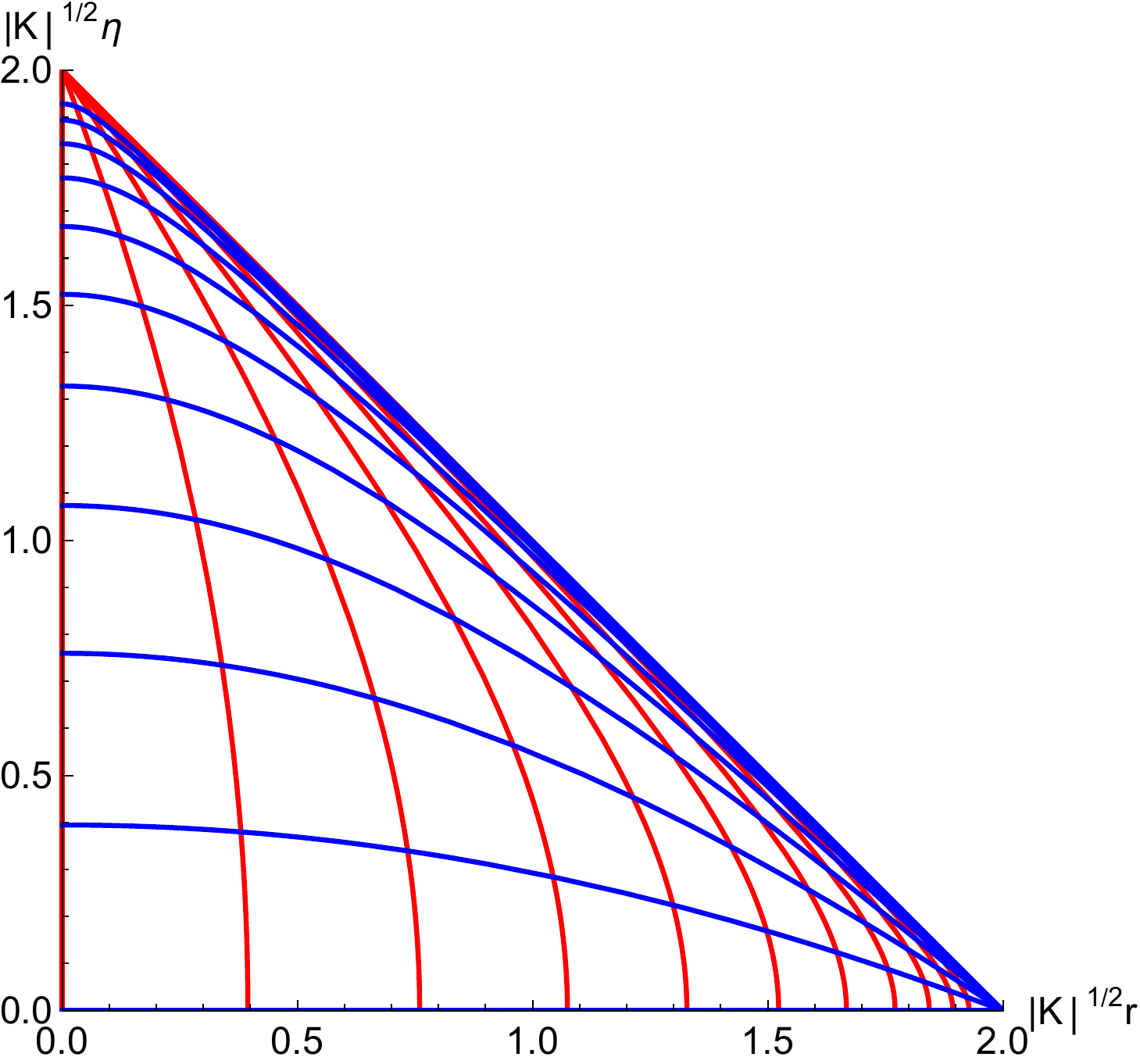}
		\end{center}
		\caption{\small Conformally flat coordinates in a closed FRW universe, $K>0$ (left panel), and in an open FRW universe, $K<0$ (right panel). The red curves represent $\chi = \text{const.}$ surfaces, the blue ones are $\tau = \text{const.}$ surfaces. In the left panel, the red curves are for $\abs{K}^{1/2}\chi = 0.008, 0.2,0.6,1.2,1.8,2.4$ from left to right, and the blue curves are for the same values of $\abs{K}^{1/2}(\tau-\tau_0)$ from bottom to top. In the right panel, the red curves are for $\abs{K}^{1/2}\chi = 0, 0.4,0.8,\ldots,3.6,4.0$ from left to right, and the blue curves are for the same values of $\abs{K}^{1/2}(\tau-\tau_0)$ from bottom to top. In both plots, the curves can be extended to negative values of $\eta$ by reflecting about the $x-$axis.
		}\label{FRWcoord}
\end{figure}

The general Robertson-Walker (RW) line element in spherical coordinates reads:
\begin{equation}
		ds^2 = -dt^2 + a^2(t) \left(\frac{d\bar{r}^2}{1-K \bar{r}^2} + \bar{r}^2 d\Omega^2\right),
\end{equation}
where $d\Omega^2$ is the unit two-sphere metric and $K$ is the Gaussian curvature of the spatial sections. Here we adopt the convention that $a$ is scaled to unity at the present time, whereby $\bar{r}$ has the dimension of distance and $K$ inverse distance squared; curvature $K$ can take any real value. With the coordinate transformation
\begin{equation}\label{SK}
		\bar{r} = S_K(\chi) \ , \quad S_K(\chi) = \begin{cases}
			K^{-1/2}\sin (K^{1/2} \chi)\ , &\quad K > 0 \\
			\chi\ , &\quad K = 0 \\
			\abs{K}^{-1/2}\sinh (\abs{K}^{1/2}\chi)\ , &\quad K < 0 \ ,
		\end{cases}
\end{equation}
where $S_K$ is the generalized sine, we get the RW metric with a new radial coordinate $\chi$:
\begin{equation}
	ds^2 = -dt^2 +a(t)^2 \left(
	d\chi^2 + S_K(\chi)^2 d\Omega^2
	\right) \ .
\end{equation}
Notice that since $a=1$ today, $\chi$ is the physical spatial distance at the present time.
We then introduce parametric\footnote{Notice that this is usually called `conformal time' but here we reserve that name for the time coordinate in a conformally flat frame.} time coordinate $\tau = \int dt/a(t)$, which brings us one step closer to the conformally flat form:
\begin{equation}
		ds^2 = a(\tau)^2 \left(-d\tau^2+
		d\chi^2 + S_K(\chi)^2 d\Omega^2
		\right) \ .
\end{equation}
Now the line element inside the brackets corresponds to the Einstein static universe ($K >0$), the Minkowski spacetime ($K=0$), or a hyperbolic space static in time ($K<0$). Carrying on, we make a final transformation to a conformally flat coordinate system by \cite{Gron:2011yi} (see also \cite{Harada:2018ikn})
\begin{equation}
		\eta = \frac{1}{2}(f(\tau + \chi) + g(\tau - \chi))\ , \quad r = \frac{1}{2}(f(\tau +\chi) - g(\tau -\chi)) \ ,
\end{equation}
where $f$ and $g$ must satisfy
\begin{align}\label{fgcond}
		\frac{1}{4}(f(\tau + \chi) - g(\tau - \chi))^2 = r^2 = f'(\tau + \chi) g'(\tau - \chi) S_K(\chi)^2  \ .
\end{align}
The metric then reads
\begin{equation}
		ds^2 = \frac{a(\tau)^2}{f'(\tau + \chi)g'(\tau - \chi)} \left( 
		- d\eta^2 + dr^2 + r^2 d\Omega^2
		\right) \ ,
\end{equation}
where conformal flatness is now fully manifest. From (\ref{fgcond}) we find, by setting $\chi = 0$, that we must in fact have $f = g$, whereby we get a differential equation of one unknown function
\begin{equation}\label{feq}
		f'(U)f'(V) S_K ((V-U)/2)^2 = \frac{1}{4}(f(V)-f(U))^2 \ ,
\end{equation}
where we introduced the lightcone coordinates
\begin{equation}
		U = \tau -\chi\ , \quad V = \tau + \chi \ .
\end{equation}
Notice that
\begin{equation}\label{fU}
		u = \eta - r = f(\tau-\chi)=f(U) \ ,
\end{equation}
{\emph{i.e.}}, constant $U$ null surfaces coincide with constant $u$ ones.
Setting $U = V_0 = \text{const.}$, Eq. (\ref{feq}) can be integrated to yield
\begin{equation}\label{trafosol}
		f(V) = f(V_0) + \frac{2 f'(V_0)}{C + I_K \left(\frac{V-V_0}{2}\right)}\ ,
\end{equation}
where $I_K$ is defined as
\begin{equation}
		I_K (\chi) = 
		\begin{cases}
			\sqrt{K}\cot (\sqrt{K}\chi)\ , & K>0 \\
			1/\chi \ , & K = 0 \\
			\sqrt{-K}\coth (\sqrt{-K}\chi)\ , &K<0 \ ,
		\end{cases}
\end{equation}
satisfying $I'_K(\chi) = - 1/S_K(\chi)^2$.
The overall conformal factor then is
\begin{equation}\label{Omega}
		\Omega = \frac{2\, a(\tau) S_K(\chi)}{f(V)-f(U)} = \frac{a(\tau) S_K(\chi)}{r(\tau,\chi)} \ .
\end{equation}
The solution (\ref{trafosol}) allows for a residual freedom to choose coordinates when transforming the metric to a conformally flat form. We choose a simple transformation with $V_0=\tau_0, f(V_0) = C = 0, f'(V_0) = 1$, where $\tau_0$ is the value of parametric time today, so that
\begin{equation}\label{trafof}
		f(V) = 2 I_K((V-\tau_0)/2)^{-1} = \begin{dcases}
			2 K^{-1/2} \tan (\sqrt{K}(V-\tau_0)/2) \ , &K>0 \\
			V - \tau_0 \ , & K=0 \\
			2 (-K)^{-1/2} \tanh (\sqrt{-K}(V-\tau_0)/2) \ , &K<0
	\end{dcases} \ .
\end{equation}
For an illustration of the coordinate system, see Fig. \ref{FRWcoord}. This choice of coordinates brings the conformal factor to unity at the origin $\chi = 0$ of the $\tau=\tau_0$ hypersurface.
The conformal factor now is, explicitly,
\begin{equation}
		\Omega = \begin{dcases}
		\frac{a(\tau)\sin (\sqrt{K}\chi)}{\tan(\sqrt{K}(\tau-\tau_0+\chi)/2)-\tan(\sqrt{K}(\tau-\tau_0-\chi)/2)}  \ , &K>0 \\
		a(\tau) \ , & K=0 \\
		\frac{a(\tau)\sinh (\sqrt{-K}\chi)}{\tanh(\sqrt{-K}(\tau-\tau_0+\chi)/2)-\tanh(\sqrt{-K}(\tau-\tau_0-\chi)/2)} \ , &K<0
		\end{dcases} \ ,
\end{equation}
and restricting to the $\tau=\tau_0$ spacelike hypersurface, denoted by $\Sigma_0$, it simplifies to
\begin{equation}
		\Omega\big\rvert_{\Sigma_0} = \begin{cases}
			\cos^2 (\sqrt{K}\chi/2)  \ , &K>0 \\
			1 \ , & K=0 \\
			\cosh^2 (\sqrt{-K}\chi/2) \ , &K<0
		\end{cases} \ ,
\end{equation}
where we used the convention that $a = 1$ at the present time. This is a rather convenient simplification because we are primarily interested in evaluating the memory effect today, and in all spatial geometries $\Omega\big\rvert_{\Sigma_0}$ is close to unity when $\sqrt{\abs{K}}\chi \ll 1$. Furthermore, this choice of coordinates is advantageous in view of the large $r$ expansions \nr{asympt1}--\nr{asympt2}, as large $r$ now corresponds to a large physical distance via $a(\tau_0)\chi = \chi \approx S_K(\chi) \approx r$, making the expansions meaningful. Here `large' of course means large compared to the source length scales, but not large compared to the Hubble distance or the radius of curvature of the universe.
		
\subsection{Memory effect in a spatially curved FRW spacetime}

\subsubsection{Projection to local observer frame}
	
Now that we have found a conformally flat coordinate system for the FRW metric, we need to relate our solution for the memory tensor to the persisting displacement measured in the local inertial frame of the comoving observer at $(t_0,\chi,\theta,\phi)$. Recall that in \nr{memeffect} the memory effect was given by
\begin{equation}
	\Delta s^a \approx s^b \int_{t_i}^{t_f}dt \int_{t_i}^{t}dt' \,
    \tensor{C}{^a_t_t_b} \ ,
\end{equation}
where $\tensor{C}{^a_t_t_b}$ is the Weyl tensor projected to the local observer frame:
\begin{equation}
    \tensor{C}{^a_t_t_b} = e^a\mud \tensor{C}{^\mu_\nu_\rho_\sigma} e\nup_t e^\rho_t e^\sigma_b \ .
\end{equation}
A suitable orthonormal basis $\{ e\mup_t, e\mup_a \}$ is given in the comoving coordinates $(t,\chi,\theta,\phi)$ by
\begin{align}
    &e\mup_t = u\mup = (1,0,0,0) \ ,  &&e\mup_1 = (0,a(t)\pminus{1}, 0,0) \ , \nn \\
    &e\mup_2 = (0,0,a(t)\pminus{1}S_K(\chi)\pminus{1},0) \ , \quad &&e\mup_3 = (0,0,0,a(t)\pminus{1}S_K(\chi)\pminus{1}(\sin\theta)\pminus{1}) \ .
\end{align}
One can easily check that these all are parallel transported along the observer worldline. Here the spatial vielbein is aligned with the spherical coordinate frame evaluated at the observer position. Assuming that the detector system lies in the plane transversal to the radial direction, we can write the displacement with the angular coordinates of the comoving frame:
\begin{equation}
	\Delta s^A \approx s^B \int_{t_i}^{t_f}dt \int_{t_i}^{t}dt' \,
    \tensor{C}{^A_t_t_B} \ .
\end{equation}
We now relate this to the memory tensor.
For this, we need to take into account that in the equations \nr{final1}--\nr{final2} for the memory tensor, we have quantities that are defined by integrals over conformal observer time. We should express these, too, in terms of proper time measured by the observer. We will evaluate these expressions at the observer position at the present time assuming that the burst duration $\delta \tau$ is very small compared to the Hubble time or the curvature radius of the universe.\footnote{Concretely, we could, {\emph{e.g.}}, have $\delta\tau = 10^3\, \text{yr}$ and $H_0\pminus{1} \sim 10^{10} \,\text{yr}$ so that $\sqrt{\abs{K}}\delta\tau < H_0 \delta\tau \sim 10\pminus{7}$.}
Keeping in mind the relation between comoving and parametric time $dt/d\tau = a$, and assuming that the scale factor does not vary significantly over the observer timescale, we may approximate
\bea
		\int_{t_i}^{t_f} dt\int_{t_i}^{t}dt' \tensor{C}{^A_t_t_B}(t') &\approx & \int_{\tau_i}^{\tau_f} d\tau \int_{\tau_i}^{\tau}d\tau' \tensor{C}{^A_\tau_\tau_B}(\tau') \nn \\ 
		& \approx & a(\tau_0)^{-2} S_K(\chi)^{-2} h^{AC} \int_{\tau_i}^{\tau_f} d\tau \int_{\tau_i}^{\tau}d\tau' \tensor{C}{_C_\tau_\tau_B}(\tau')  \ . \label{jacobirhs}
\eea
Then we just need to relate this to the Weyl tensor components in the conformally flat frame. Since the Jacobian determinant of the transformation from $\tau$ to $U = \tau - \chi$ is unity, we get
\bea
		\int_{\tau_i}^{\tau_f} d\tau \int_{\tau_i}^{\tau}d\tau' \tensor{C}{_A_\tau_\tau_B}
		& = & \int_{U_i}^{U_f} dU \int_{U_i}^{U}dU' \tensor{C}{_A_\tau_\tau_B} \nn \\
		& = & \int_{u_i}^{u_f} du \, (f^{-1})'(u) \int_{u_i}^{u}du' (f^{-1})'(u') \tensor{C}{_A_\tau_\tau_B}  \ ,
\eea
where we used the relation \nr{fU}. We expand the derivative of the inverse coordinate transformation:
\begin{equation}\label{dfinv}
		(f^{-1})'(u) = \frac{1}{1 + \frac{1}{4}K u^2} = 1-\frac{1}{4}K \chi^2 + \order{K\chi\delta\tau,(K \chi^2)^2} \ ,
\end{equation}
where we have small correction terms proportional to the burst duration.
Then, we transform the tensor components:
\bea
		\tensor{C}{_A_\tau_\tau_B} & = & \tensor{C}{_A_\mu_\nu_B} \frac{\pt (\eta,r)\mup}{\pt\tau}\frac{\pt (\eta,r)\nup}{\pt\tau} \nn \\ 
		& = & \tensor{C}{_A_\eta_\eta_B}\left(\frac{\pt \eta}{\pt\tau}\right)^2 + \tensor{C}{_A_r_r_B}\left(\frac{\pt r}{\pt\tau}\right)^2 + \left(\tensor{C}{_A_\eta_r_B}+\tensor{C}{_A_r_\eta_B}\right)\frac{\pt\eta}{\pt\tau}\frac{\pt r}{\pt\tau} \ . \label{Ctrafo}
\eea
Expanding also the Jacobian matrix elements, we get
\bea
		\frac{\pt \eta}{\pt\tau} & = & 1 + \frac{1}{4}K\chi^2 + \mathcal{O}(K \delta\tau^2, (K\chi^2)^2)  \\
		\frac{\pt r}{\pt\tau}  & = & \frac{1}{2}K\chi\delta\tau + \mathcal{O}(K^2 \chi^3\delta\tau) \ .
\eea
Hence we see that in \nr{Ctrafo} the last two terms are negligible, and we are only left with
\begin{equation}\label{Cappro}
		\tensor{C}{_A_\tau_\tau_B} = (1+\frac{1}{2}K\chi^2)C_{A\eta\eta B} + \order{K\chi\delta\tau,(K\chi^2)^2} \ .
\end{equation}
The leading $K$ corrections from \nr{dfinv} and \nr{Cappro} cancel out so that we have
\begin{equation}
	\int_{\tau_i}^{\tau_f} d\tau \int_{\tau_i}^{\tau}d\tau' \tensor{C}{_A_\tau_\tau_B} = \int_{u_i}^{u_f} du  \int_{u_i}^{u}du' \tensor{C}{_A_\eta_\eta_B} + \order{K\chi\delta\tau,(K\chi^2)^2} \ .
\end{equation}
The RHS correction terms are very small for our system so they can be neglected. We then want to relate the RHS double integral to the memory tensor \nr{memorydef}.
From tracelessness of the Weyl tensor it follows that $\tensor{E}{_C^C} \sim r^{-1}$ so that, asymptotically,
\begin{equation}
e_{AB} \approx r^{-1}\tilde{E}_{AB} = r^{-1}\left(E_{AB}-\frac{1}{2}h_{AB} \tensor{E}{_C^C}\right) \approx r^{-1} E_{AB} \ .
\end{equation}
Hence,
\begin{equation}
	C_{A\eta\eta B} = - \Omega E_{AB} \approx - \Omega\, r\, e_{AB} = - a(\tau) S_K(\chi) e\ABD \ .
\end{equation}
Plugging this in the integral, we obtain
\begin{equation}
		\int_{u_i}^{u_f}du \int_{u_i}^{u} du'  C_{A\eta\eta B} \approx -\int_{u_i}^{u_f}du \int_{u_i}^{u} du' a(\tau_0)S_K(\chi) e\ABD \approx a(\tau_0)S_K(\chi)m\ABD
\end{equation}
so that
\begin{equation}
		\int_{t_i}^{t_f} dt\int_{t_i}^{t}dt' \tensor{C}{^A_t_t_B}(t') \approx -\frac{\tensor{m}{^A_B}}{S_K(\chi)} \ ,
\end{equation}
where we used \nr{jacobirhs} and raised the $A$ index with the unit two-sphere metric. The memory effect is then given by
\begin{equation}\label{memformula1}
		\Delta s^A \approx -\frac{\tensor{m}{^A_B} s^B}{S_K(\chi)} \ ,
\end{equation}
where the memory tensor $\tensor{m}{^A_B}$ is determined by \nr{memorysol1}.

\subsubsection{Background curvature corrections to the flat space memory effect}

The memory effect consists of two pieces that are called ordinary and null memory. To emphasize this distinction, we write 
\begin{equation}\label{OandN}
    \tensor{m}{^A_B}=\tensor{(m_O)}{^A_B} + \tensor{(m_N)}{^A_B} \ ,
\end{equation}
where $m_O$ gives rise to the ordinary memory and $m_N$ to null memory. In typical situations, the null piece is more interesting, and we will focus on it.
Recall that in the solution \nr{memorysol1}--\nr{memorysol2} for the memory tensor there is a background dependent contribution to null memory, which makes \nr{memformula1} different from the corresponding results in Minkowski space and flat FRW spacetime.
We would like to express the final result in terms of the flat space memory tensor and background curvature corrections stemming from the factor involving $\Omega$ and its derivatives on the RHS of \nr{memorysol2}.  The term on the RHS of \nr{memorysol2} in question reads
\begin{equation}\label{Omegafac}
		\Omega^{-2}(\Omega + r\pt_r \Omega + r \pt_\eta \Omega) = \frac{r}{\Omega}\frac{\pt_r(\Omega r)+\pt_\eta(\Omega r)}{\Omega r} = \frac{r}{\Omega} \left( \frac{a'}{a}(\pt_\eta\tau + \pt_r \tau) + \frac{S'_K}{S_K}(\pt_\eta\chi+ \pt_r \chi) \right) \ ,
\end{equation}
where we used \nr{Omega}.
The nontrivial part of the inverse coordinate transformation Jacobian matrix is
\begin{equation}
		J^{-1} = \begin{pmatrix}
			\dfrac{\pt\tau}{\pt\eta} & \dfrac{\pt\tau}{\pt r} \\
			\dfrac{\pt\chi}{\pt\eta} & \dfrac{\pt\chi}{\pt r}
		\end{pmatrix} 
		= \frac{1}{2}\begin{pmatrix}
		\dfrac{f'(\tau-\chi)+f'(\tau+\chi)}{f'(\tau-\chi)f'(\tau+\chi)} & \dfrac{f'(\tau-\chi)-f'(\tau+\chi)}{f'(\tau-\chi)f'(\tau+\chi)} \\
		\dfrac{f'(\tau-\chi)-f'(\tau+\chi)}{f'(\tau-\chi)f'(\tau+\chi)} & \dfrac{f'(\tau-\chi)+f'(\tau+\chi)}{f'(\tau-\chi)f'(\tau+\chi)}
		\end{pmatrix} \ .
\end{equation}
Hence, \nr{Omegafac} becomes
\bea
		\frac{\pt_r(\Omega r)+\pt_\eta(\Omega r)}{\Omega^2} 
		& = & \frac{r}{\Omega}\frac{1}{f'(\tau+\chi)}\left(\frac{a'}{a} + \frac{S'_K}{S_K}\right) \nn \\
            & = & \frac{r}{\Omega}\frac{1}{f'(\tau+\chi)S_K(\chi)}\left(\frac{a'(\tau)}{a(\tau)}S_K(\chi) + S'_K(\chi)\right) \ .
\eea
With an analysis similar to what was done for the Weyl tensor integral above, one can see that the above conformal factor term stays essentially constant over the burst, and it yields small curvature corrections to the memory tensor in \nr{memorysol2}. We should note, however, that also radiated power $L$ and, consequently, radiated energy $F$ should be related to the corresponding quantities in a comoving observer frame. This also yields corrections that need to be taken care of. To start with, write luminosity in two different coordinate systems as
\begin{equation}\label{confcom}
		T^\rmi{conf}\munud = \frac{L_\rmi{conf}}{\Omega^2 r^2}\pt\mud u \,\pt\nud u= \frac{L_\rmi{conf}}{a^2 S_K^2}\pt\mud u \, \pt\nud u \ , \quad T^\rmi{com}\munud = \frac{L_\rmi{com}}{a^2 S_K^2}\pt\mud U \pt\nud U \ ,
\end{equation}
where $T^\rmi{conf}\munud$ is our stress-energy source in the conformal coordinates (now $L^\rmi{conf}$ stands for our earlier $L$), and $T^\rmi{com}\munud$ is the source in the comoving coordinates $(\tau,\chi,\theta,\phi)$.
Solving the stress-energy conservation equation in the $(\tau,\chi,\theta,\phi)$ coordinates yields the latter one of these equalities in a similar way it did for the first one in \nr{Asol}. The stress-energy tensor components are related simply by a coordinate transformation, which gives us one luminosity function in terms of the other one. For instance, transforming the $\eta\eta$ component yields
\be
		T^\rmi{conf}_{\eta\eta} = T^\rmi{com}\munud \frac{\pt(\tau,\chi)\mup}{\pt\eta} \frac{\pt(\tau,\chi)\nup}{\pt\eta} = T^\rmi{com}_{\tau\tau} \left(\frac{\pt\tau}{\pt\eta}\right)^2 + T^\rmi{com}_{\chi\chi} \left(\frac{\pt\chi}{\pt\eta}\right)^2 + 2 T^\rmi{com}_{\tau\chi}\frac{\pt\tau}{\pt\eta}\frac{\pt\chi}{\pt\eta} \ .
\ee
But from \nr{confcom} we see that the different components are related to each other just by a factor of $\pm 1$ so the above equation becomes
\begin{equation}
		T^\rmi{conf}_{\eta\eta} = T^\rmi{com}_{\tau\tau} \left(\frac{\pt\tau}{\pt\eta}- \frac{\pt\chi}{\pt\eta}\right)^2 = \frac{1}{f'(\tau-\chi)^2} T^\rmi{com}_{\tau\tau} \ .
\end{equation}
This gives the sought-after relation between the luminosity functions:
\begin{equation}
		L_\rmi{conf}= f'(\tau-\chi)\pminus{2}L_\rmi{com} = f'(\tau-\chi)\pminus{2}(1+z)\pminus{2}L_s \ ,
\end{equation}
where $z$ is the cosmological redshift and $L_s$ is the luminosity in the source frame. Hence we see that there is a small curvature-dependent correction, encoded in the transformation function $f$, when we express the conformal frame luminosity in terms of the comoving frame luminosity and, moreover, the result is redshifted when we express this in terms of the source frame luminosity. On the RHS of \nr{memorysol2}, we have the total radiated energy per unit solid angle, $F$, which we get by integrating luminosity over conformal time. Therefore, we have
\begin{equation}
		F_\rmi{conf} = \int du \,L_\rmi{conf} \approx f'(\tau_0-\chi)\pminus{2}(1+z)\pminus{2} \int dU f'(U) L_s \approx f'(\tau_0-\chi)\pminus{1}(1+z)\pminus{1} \int dt_s  L_s  \ ,
\end{equation}
where $t_s$ is time in the source frame. Here we again used our assumption of short burst, and we also used $d\tau = dt_s/a(t_s)$ obtaining an extra factor of $1+z$.
Denoting the total radiated energy per unit solid angle measured in the source frame by $F_s$, we see that
\begin{equation}\label{Fconf}
	F_\rmi{conf} \approx f'(\tau_0-\chi)\pminus{1}(1+z)\pminus{1} F_s \ .
\end{equation}
When we count in all relevant background curvature corrections, we get that the flat space memory tensor is multiplied by the factor
\begin{equation}
		(1+z)\pminus{1}\frac{\pt_r(\Omega r)+\pt_\eta(\Omega r)}{f'(\tau_0-\chi)\Omega^2} = \frac{r}{\Omega}\frac{(1+z)\pminus{1}}{f'(\tau_0-\chi)f'(\tau_0+\chi)S_K(\chi)}\left(\frac{a'(\tau)}{a(\tau)}S_K(\chi) + S'_K(\chi)\right) \ .
\end{equation}
Using \nr{feq}, we see that at the present time the RHS is equal to
\begin{equation}
	(1+z)\pminus{1}(a'(\tau_0)S_K(\chi) + C_K(\chi)) \ ,
\end{equation}
where we have defined
\begin{equation}
		C_K(\chi)=
		\begin{cases}
		\cos (K^{1/2} \chi)\ , &\quad K > 0 \\
		1\ , &\quad K = 0 \\
		\cosh (\abs{K}^{1/2}\chi)\ , &\quad K < 0
		\end{cases}  \ .
\end{equation}
Notice that this can be written in terms of the Hubble constant, redshift, and luminosity distance $d_L$ as
\begin{equation}
		(1+z)\pminus{1}\left(\frac{H d_L}{1+z} + C_K(\chi)\right) \ ,
\end{equation}
recalling that the luminosity distance is defined by
\begin{equation}
		d_L = a(\tau_0)(1+z) \bar{r} = (1+z) S_K(\chi) \ .
\end{equation}
So when we express the radial function $S_K$ in terms of redshift and luminosity distance, 
the equation \nr{memformula1} for the null memory effect becomes
\begin{equation}\label{memformula2}
    \Delta s^A \approx - \frac{(1+z)\pminus{1}H d_L + C_K(\chi)}{d_L} \tensor{(m^f_N)}{^A_B} s^B \ ,
\end{equation}
where $\chi$ is the physical distance to the source at present time, and $(m^f_N)\ABD$ is the null memory tensor computed in flat background with distance in the Minkowski space matching with $d_L$ in the FRW universe. This is the second main result of our paper. Notice that when $K=0$ and the redshift is small, the numerator is approximately just $1+z$. On the other hand, with nonzero but small $K$ we can expand about $K\chi^2 = 0$ to see the first curvature corrections to the flat space memory:
\begin{equation}
		\Delta s^A \approx -\frac{1+H_0 \chi - \frac{1}{2}K\chi^2 - \frac{1}{6}H_0 K \chi^3}{d_L} \tensor{(m^f_N)}{^A_B} s^B \ .
\end{equation}
For practical purposes, it might be useful to express this also in terms of the curvature energy density parameter $\Omega_K = - K/H_0^2$:
\begin{equation}
	\Delta s^A \approx -\frac{1+H_0 \chi + \frac{1}{2}\Omega_K(H_0 \chi)^2 + \frac{1}{6}\Omega_K (H_0\chi)^3}{d_L} \tensor{(m^f_N)}{^A_B} s^B \ .
\end{equation}
Notice that $1+H_0 \chi$ is the factor that was found in \cite{Bieri:2015jwa}. For physical distance $\chi$ comparable to the Hubble length, the curvature terms might also become important depending on the magnitude of $\Omega_K$. Null memory is enhanced by negative curvature, whereas positive curvature dampens it. Ordinary memory, on the other hand, is simply enhanced by a redshift factor when expressed in terms of the luminosity distance.

\section{Discussion}

In this paper, we generalized the Weyl tensor analysis \cite{Bieri:2013ada,Bieri:2015jwa} of the memory effect to a generic conformally flat background geometry. The advantage of the method was that we were able to directly handle gauge-invariant physical quantities and bypass the question of gauge-fixing and physical interpretation thereof. Conformally flat background geometry allowed us to study the memory via Weyl tensor perturbations, which are manifestly physical. In this approach, both ordinary and null memory can be studied straightforwardly. However, gauge-invariance of the method only holds to first order in perturbation theory but is broken at higher orders, which is why the analysis does not capture, {\emph{e.g.}}, Christodoulou's nonlinear memory effect \cite{Christodoulou:1991cr}, as was already pointed out in \cite{Bieri:2013ada}. Extension of the method to higher perturbative orders still remains an open problem. Asymptotic symmetry structure at null infinity might offer useful guidance in this vein. Of course, the analysis of symmetry transformations in general conformally flat spacetimes is interesting in its own right.
	
In our analysis, we assumed certain fall-off conditions \nr{asympt1}--\nr{asympt2} for the Weyl tensor components. These same fall-offs were justified and used in the Minkowski background in \cite{Bieri:2013ada}, and in \cite{Bieri:2015jwa} they were assumed to apply also in de Sitter background. Imposing these fall-off conditions seems reasonable in a curved background, too, as long as the background curvature varies over much larger scales than the curvature perturbations. Our choice of coordinates ensured that large coordinate distance corresponds to large physical distance at present time, which indicates that the expansions at least were meaningful. However, this argument falls short of a full-blown proof and it would be nice to have a more sound justification. We leave further analysis of the asymptotics for future work.
	
We also assumed that the GW burst is short compared to background length scales. This assumption of course only applies to, for instance, binary systems with short lifetimes, or situations where we only consider the final stage of the binary inspiral. In general, however, binary lifetimes can be anything up to the Hubble time \cite{vanSon:2021zpk}. It would be interesting to generalize our calculation to cover also long-lived processes.
 
We applied the generic solutions to the special case of a spatially curved FRW geometry. Other potentially interesting applications of this approach include the Stephani universe, which is a conformally flat inhomogeneous cosmological model (see, \emph{e.g.}, \cite{Stephani:2003tm}), and $AdS_n\times {\mathbb{S}}^n$ (in particular, because $AdS_2\times{\mathbb{S}}^2$ is the extremal near-horion limit of the Reissner-Nordstr\"om black hole). Applications in dimension $n > 4$, however, would require a dimensional generalization of the perturbative analysis performed in Sec. \ref{weylsec}.

In the FRW spacetime, in the zero curvature $K=0$, small redshift, limit we retrieve the previously reported results that the memory effect is enhanced by a redshift factor \cite{Bieri:2015jwa,Bieri:2017vni,Tolish:2016ggo,Chu:2016ngc,Jokela:2022rhk}. With a non-zero curvature and sizable redshift of the source event, our main result predicts that there is a curvature- and distance-dependent correction to the memory effect, compared to the flat FRW case. Based on the stringent observational bounds on the curvature density parameter \cite{Planck:2018vyg}, current GW measurements are far too inaccurate to capture such small corrections. Indeed, it is difficult to determine the presence or absence of the memory effect with the current detectors in the first place, let alone small corrections thereof. However, it might not be too far-fetched to expect that in the future the detectors will be vastly improved in accuracy, maybe even to the extent that we shall have precise measurements of the memory effect on a regular basis. Were that to be the case, we would potentially have a new, independent way of constraining the curvature density parameter, thus making the memory effect a novel source of data on the large-scale structure of our universe. This is a rather intriguing prospect, albeit quite a distant one for now.

\paragraph{Acknowledgements}
We want to thank David Garfinkle and Jan W. van Holten for useful comments and correspondence. N.~J. and M.~S. have been supported in part by the Academy of Finland grant no. 1322307. M.~S. is also supported by the Finnish Cultural Foundation.

\appendix
\section{Weyl tensor asymptotics}\label{app:weyl}
Here we give an argument for the asymptotic behavior of the Weyl tensor stated in \nr{asympt1}--\nr{asympt2}. We show that in the high frequency limit, the Cartesian components $\Omega\pminus{1}C_{\rho\sigma\mu\nu}$ fall off as $r\pminus{1}$, from which the conditions \nr{asympt1}--\nr{asympt2} for the spherical coordinate components follow by standard coordinate transformation rules and analysis of constraint equations \nr{Econstraint}--\nr{Bconstraint}.

We start by deriving a nonlinear wave equation for the Riemann tensor. The differential Bianchi identity is
\begin{equation}\label{bianchi2}
    \na_{[\epsilon}R_{\rho\sigma]\mu\nu} = 0 \ .
\end{equation}
Contracting this with $g^{\epsilon\nu}$ yields
\begin{equation}\label{bianchi3}
    \na^\nu R_{\rho\sigma\mu\nu} = \na_\sigma R_{\rho\mu} - \na_\rho R_{\sigma\mu} \ ,
\end{equation}
after which we contract \nr{bianchi2} with $\na^\epsilon$ and use \nr{bianchi3} to get the wave equation for the Riemann tensor \cite{vanHolten:2022zsw}:
\begin{equation}\label{riemannwave}
    \na^\eps\na_\eps R_{\rho\sigma\mu\nu} - 2 \tensor{R}{^\eps_\rho_\sigma^\gamma} R_{\eps\gamma\mu\nu} + 2 R_{\eps\rho\mu\gamma} \tensor{R}{^\eps_\sigma_\nu^\gamma} - 2 \tensor{R}{^\eps_\rho_\nu^\gamma} R_{\eps\sigma\mu\gamma}  + 2 \tensor{R}{_{[\rho}^\eps} \tensor{R}{_{\sigma]}_\eps_\mu_\nu} = 2 \na_\rho \na_{[\mu}R_{\nu]\sigma} - 2 \na_\sigma \na_{[\mu}R_{\nu]\rho} \ .
\end{equation}
Next we perturb this wave equation to linear order, and apply the high frequency approximation \cite{Maggiore:2007ulw} where the metric perturbations $h\munud$, the typical perturbation wavelength $\lambda$, and the background curvature radius $L_B$ satisfy 
\begin{equation}\label{separation}
    \abs{h\munud} \equiv h \ll \frac{\lambda}{L_B} \ll 1 \ .
\end{equation}
We will only include leading order and next-to-leading order (NLO) terms in our approximation. The background metric can be brought to $\order{1}$ by a coordinate rescaling, the background Christoffel symbols are $\bar{\Gamma}^\rho\munud =\order{L_B\pminus{1}}$, and the background curvature tensors are order $\order{L_B\pminus{2}}$. First derivatives of the metric perturbation are $\pt h\munud = \order{h/\lambda}$, and the second derivatives $\pt^2 h\munud = \order{h/\lambda^2}$. Perturbation of the first term in \nr{riemannwave} is, schematically,
\begin{equation}\label{boxriem}
    \Box \, \delta R + \delta g\pminus{1} \pt^2 \bar{R} - \bar{\Gamma}\pt\,\delta R - \delta \Gamma \pt \bar{R} + \delta \Gamma \, \bar{\Gamma} \bar{R} + \bar{\Gamma}\bar{\Gamma} \delta R \ ,
\end{equation}
where $\Box \equiv \pt^\eps \pt_\eps$, and we dropped all indices and clumped all terms of the same type into one for brevity's sake.
Curvature perturbations are $\delta R = \order{h/\lambda^2}$. The leading order contribution in \nr{riemannwave} hence comes from $\Box\, \delta R = \order{h/\lambda^4}$. The third term in \nr{boxriem} is NLO:
\begin{equation}
    \bar{\Gamma} \pt \, \delta R \sim \frac{1}{L_B} \frac{h}{\lambda^3} \sim \frac{\lambda}{L_B} (\Box \,\delta R) \ ,
\end{equation}
which we need to include in our analysis. The rest of terms in \nr{boxriem}, however, are next-to-next-to-leading order (NNLO) or smaller:
\begin{equation}
    \bar{\Gamma}\bar{\Gamma} \delta R \sim \frac{1}{L_B^2} \frac{h}{\lambda^2} \sim \left(\frac{\lambda}{L_B}\right)^2 (\Box \, \delta R) \ ,
\end{equation}
and the rest three are even more strongly suppressed by $\lambda/L_B$. Also, notice that all the quadratic curvature terms in \nr{riemannwave} are negligible; their perturbations are, schematically, either
\begin{equation}
    \delta g\pminus{1} \bar{R}\bar{R} \sim \frac{h}{L_B^4} \sim (\Box \, \delta R) \left(\frac{\lambda}{L_B}\right)^4 \ , \quad 
\end{equation}
or
\begin{equation}
    \delta R \,\bar{R} \sim \frac{h}{\lambda^2}\frac{1}{L_B^2} \sim (\Box \, \delta R) \left(\frac{\lambda}{L_B}\right)^2 \ .
\end{equation}
Thus, the only significant contributions on the LHS of \nr{riemannwave} come from the first term. Before we do further analysis on these, we turn to the RHS of \nr{riemannwave}. Here also the leading contributions come from $\pt^2 \, \delta R$ terms, NLO terms are $\bar{\Gamma}\pt \,\delta R$, and the rest are NNLO or smaller. With Einstein's equation, the RHS can be written in terms of stress-energy perturbations; we denote the LO and NLO source terms collectively as $S_{\rho\sigma\mu\nu}$.

We now write the significant perturbative contributions from \nr{riemannwave} LHS first term in detail:
\begin{align}\label{boxpert}
    &\delta (\na\au \na\ad R_{\rho\sigma\mu\nu}) \nn \\
    \simeq &\bar{g}\abu \pt_\alpha \pt_\beta \, \delta R_{\rho\sigma\mu\nu} - \bar{g}\abu \bar{\Gamma}^\gamma\abd \pt_\gamma \, \delta R_{\rho\sigma\mu\nu} 
    \nn \\
    &- 2 \bar{g}\abu (\bar{\Gamma}^\gamma_{\alpha\rho}\pt_\beta  \delta R_{\gamma\sigma\mu\nu} + \bar{\Gamma}^\gamma_{\alpha\sigma}\pt_\beta  \delta R_{\rho\gamma\mu\nu} + \bar{\Gamma}^\gamma_{\alpha\mu}\pt_\beta  \delta R_{\rho\sigma\gamma\nu} + \bar{\Gamma}^\gamma_{\alpha\nu} \pt_\beta \delta R_{\rho\sigma\mu\gamma} ) \nn \\
    =&  \Omega\pminus{2} \pt^\alpha \pt_\alpha \, \delta R_{\rho\sigma\mu\nu} + 2 \Omega\pminus{3}\pt\ad \Omega \,\pt\au \delta R_{\rho\sigma\mu\nu}
    \nn \\
    &- 2 \Omega\pminus{3}(\pt_\rho \Omega\, \pt^\gamma \,\delta R_{\gamma\sigma\mu\nu} + \pt^\gamma \Omega \,\pt_\gamma \delta R_{\rho\sigma\mu\nu} - \pt^\gamma \Omega \,\pt_\rho \delta R_{\gamma\sigma\mu\nu} \nn \\
    &\hspace{1cm}+ \pt_\sigma \Omega\, \pt^\gamma \,\delta R_{\rho\gamma\mu\nu} + \pt^\gamma \Omega \,\pt_\gamma \delta R_{\rho\sigma\mu\nu} - \pt^\gamma \Omega \,\pt_\sigma \delta R_{\rho\gamma\mu\nu} \nn \\
    &\hspace{1cm}+ \pt_\mu \Omega\, \pt^\gamma \,\delta R_{\rho\sigma\gamma\nu} + \pt^\gamma \Omega \,\pt_\gamma \delta R_{\rho\sigma\mu\nu} - \pt^\gamma \Omega \,\pt_\mu \delta R_{\rho\sigma\gamma\nu} \nn \\
    &\hspace{1cm}+ \pt_\nu \Omega\, \pt^\gamma \,\delta R_{\rho\sigma\mu\gamma} + \pt^\gamma \Omega \,\pt_\gamma \delta R_{\rho\sigma\mu\nu} - \pt^\gamma \Omega \,\pt_\nu \delta R_{\rho\sigma\mu\gamma})
\end{align}
where we now raised indices with the Minkowski metric and used \nr{cartchris} for the background Christoffel symbols. To simplify this, we then perturb the Bianchi identity
\begin{equation}
    0 = \delta (\na_\eps R_{\rho\sigma\mu\nu} + \na_\rho R_{\sigma\eps\mu\nu} + \na_\sigma R_{\eps\rho\mu\nu} ) = \pt_\eps \, \delta R_{\rho\sigma\mu\nu} + \pt_\rho \, \delta R_{\sigma\eps\mu\nu} + \pt_\sigma \, \delta R_{\eps\rho\mu\nu} + \order{\frac{\lambda}{L_B}\pt \, \delta R} \ ,
\end{equation}
and the contracted Bianchi identity
\begin{equation}
    \pt\nup \delta R_{\rho\sigma\mu\nu} + \order{\frac{\lambda}{L_B}\pt \, \delta R} = \delta (\na\nup R_{\rho\sigma\mu\nu}) = \delta (\na_\sigma R_{\rho\mu} - \na_\rho R_{\sigma\mu}) = \pt_\sigma \, \delta R_{\rho\mu} - \pt_\rho\, \delta R_{\sigma\mu} + \order{\frac{\lambda}{L_B}\pt \, \delta R} \ .
\end{equation}
With the latter identity, we can translate the divergence terms inside the brackets in \nr{boxpert} into stress-energy source terms and negligible NNLO terms (using the fact that $\pt\Omega \sim L_B\pminus{1}$). Then, with the former identity and Riemann tensor symmetries, one can use the terms with a negative sign inside the brackets to cancel two of the $\pt^\gamma \Omega \,\pt_\gamma \delta R_{\rho\sigma\mu\nu}$ terms.
Eq. \nr{boxpert} thus simplifies to
\begin{equation}
    \delta (\na\au \na\ad R_{\rho\sigma\mu\nu}) \simeq \Omega\pminus{2}\eta\abu \pt_\alpha \pt_\beta \, \delta R_{\rho\sigma\mu\nu} - 2 \Omega\pminus{3}\pt\ad \Omega \,\pt\au \delta R_{\rho\sigma\mu\nu} + \text{source terms} \ .
\end{equation}
Next, we move the source terms to the RHS of \nr{riemannwave}, multiply both sides by $\Omega$, and rename the new RHS source as $\tilde{S}_{\rho\sigma\mu\nu}$; the equation becomes
\begin{equation}\label{almostthere}
    \Omega\pminus{1} \Box \,\delta R_{\rho\sigma\mu\nu} - 2 \Omega\pminus{2}\pt\ad \Omega \,\pt\au \delta R_{\rho\sigma\mu\nu} \simeq \tilde{S}_{\rho\sigma\mu\nu} \ .
\end{equation}
Now we manipulate the first term above a bit. Writing
\bea
    \Omega\pminus{1} \Box \,\delta R_{\rho\sigma\mu\nu} & = & \Box (\Omega\pminus{1} \delta R_{\rho\sigma\mu\nu}) + 2 \Omega\pminus{2}\pt\ad \Omega \, \pt\au \delta R_{\rho\sigma\mu\nu} + \order{\pt^2 \Omega \, \delta R} \nn \\
    & = & \Box (\Omega\pminus{1} \delta R_{\rho\sigma\mu\nu}) + 2 \Omega\pminus{2}\pt\ad \Omega \, \pt\au \delta R_{\rho\sigma\mu\nu} + \order{\left(\frac{\lambda}{L_B}\right)^2 \pt^2 \, \delta R} \ .
\eea
Hence we see that \nr{almostthere} can be recast as a flat space wave equation:
\begin{equation}
    \Box (\Omega\pminus{1} \delta R_{\rho\sigma\mu\nu}) \simeq \tilde{S}_{\rho\sigma\mu\nu} \ .
\end{equation}
With the usual Green's function $\delta(\eta-\eta' - \abs{\vec{x}-\vec{x}'})/\abs{\vec{x}-\vec{x}'}$ for the flat space box operator, we see that the rescaled Riemann perturbation must fall off as $r\pminus{1}$. On the other hand, from \nr{nullergy}--\nr{Asol} we learned that stress-energy, and hence the Ricci tensor, perturbations fall off like $r\pminus{2}$. This means that only the Weyl tensor can give the $r\pminus{1}$ behavior for the Riemann tensor perturbation. This immediately implies that the purely angular components are
\begin{equation}
    E\ABD = E_{ab} \frac{\pt x^a}{\pt x^A}\frac{\pt x^b}{\pt x^B} = \order{r} \ , \quad B\ABD = B_{ab} \frac{\pt x^a}{\pt x^A}\frac{\pt x^b}{\pt x^B} = \order{r} \ ,
\end{equation}
by a transformation from Cartesian to spherical coordinates.
Based on an analysis of constraints \nr{Econstraint}--\nr{Bconstraint} as in Appendix B of \cite{Bieri:2013ada}, with the extra assumption that $r \pt \Omega \lesssim 1$, \emph{i.e.}, $r \lesssim L_B$, we get
\begin{equation}
    E_{ra} = \order{r\pminus{2}}\ , \quad B_{ra} = \order{r\pminus{2}} \ , \quad E_{rr} = \order{r\pminus{3}} \ , \quad B_{rr} = \order{r\pminus{3}} \ .
\end{equation}
This yields the behavior \nr{asympt1}--\nr{asympt2} for the different pieces of the Weyl tensor.

\section{Solution for the memory tensor}\label{app:memsol}
In this appendix, we solve Eqs. \nr{final1} and \nr{final2} for the memory tensor $m\ABD$.
We start from the Hodge decomposition of $m\ABD$, which is a symmetric traceless tensor on the two-sphere \cite{Berger:1969}:
\begin{equation}\label{hodgedec}
		m\ABD = D_A D_B m - \frac{1}{2}h\ABD D^2 m + \frac{1}{2}\left(\eps_{AC} D^C D_B \bar{m} + \eps_{BC} D^C D_A \bar{m}\right) \ , 
\end{equation}
where $D^2$ is the Laplacian $D_A D^A$ on ${\mathbb{S}}^2$, and $m$ and $\bar{m}$ are scalar functions on the two-sphere. In general, a symmetric, traceless, and divergence-free tensor would also appear in the above decomposition but on ${\mathbb{S}}^2$, however, such a tensor vanishes identically \cite{Higuchi:1986wu}. For ease of reference we outline the proof in Appendix~\ref{app:proof}.

Taking a divergence on both sides, we get
\begin{equation}\label{div1}
    D^B m\ABD = D^B D_A D_B m - \frac{1}{2}D_A D^2 m + \frac{1}{2}\left(\eps_{AC} D^B D^C D_B \bar{m} + \eps_{BC} D^B D^C D_A \bar{m}\right) \ .
\end{equation}
Then we take another divergence to obtain an equation between scalars:
\begin{equation}\label{div2}
		D^A D^B m\ABD = D^A D^B D_A D_B m - \frac{1}{2}D^2 D^2 m + \frac{1}{2}\left(\eps_{AC}D^A D^B D_B D^C \bar{m} + \eps_{BC} D^A D^B D^C D_A \bar{m}\right) \ .
\end{equation}
By a straightforward calculation, one can then show that
\begin{equation}
		\eps_{AC}D^A D^B D_B D^C \bar{m} = \eps_{AC}D^B D^A D_B D^C \bar{m} - \eps_{AC} R^{AB} D_B D^C\bar{m} \ ,
\end{equation}
and the last term vanishes since the unit two-sphere is an Einstein manifold with $R\ABD = h\ABD$, and any scalar function $f$ satisfies $D_A D_B f = D_B D_A f$. We thus see that the two terms inside the brackets in \nr{div2} are equal. Next, one can show that
\begin{equation}
		\eps_{AC}D^B D^A D_B D^C \bar{m} = \eps^{AC} D^B \tensor{R}{_A_B_C^E}D_E \bar{m} + \eps^{AC}  \tensor{R}{_A_B_C^E} D^B D_E \bar{m} \ .
\end{equation}
The last term vanishes immediately due to symmetries. Then we use the maximally symmetric form of the Riemann tensor
\begin{equation}\label{maxsym}
    R_{ABCD} = \frac{R}{2}(h_{AC}h_{BD} - h_{AD}h_{BC})
\end{equation}
to the first term on the RHS, which yields zero. Hence, the term inside brackets in \nr{div2} vanishes, and we have
\begin{equation}\label{deltaphinew}
		D^A D^B m\ABD = D^A D^B D_A D_B m - \frac{1}{2}D^2 D^2 m \ .
\end{equation}
Then we study the first term on the RHS. By commuting the second and third covariant derivatives, one gets
\begin{equation}
		D^A D^B D_A D_B m = D^A D_A D^B D_B m + D^A (\tensor{R}{_A^C} D_C m) = D^2(D^2 + 1) m \ ,
\end{equation}
which yields
\begin{equation}
    D^A D^B m\ABD = \frac{1}{2}D^2(D^2 +2) m \ .
\end{equation}
With \nr{final2}, this gives us
\begin{equation}\label{D2phi}
		D^2 \phi = \frac{1}{2}D^2(D^2 +2) m \ ,
\end{equation}
which can be easily solved:
\begin{equation}
		\phi = \frac{1}{2}(D^2 + 2)m \ ,
\end{equation}
up to a constant shift.
Expand now both $\phi$ and $m$ in the spherical harmonic basis; from the eigenvalue equation
\begin{equation}
		D^2 Y_{lm} = - l(l+1) Y_{lm} \ ,
\end{equation}
it immediately follows that the $l = 1$ mode does not contribute to $\phi$.

Next, we will show that Eq. \nr{final2} actually entails that the part involving the Levi-Civita tensor in \nr{hodgedec} vanishes altogether. \nr{final2} implies that the RHS in \nr{div1} must be a total gradient. By commuting the first two covariant derivatives in the first term on the RHS of \nr{div1}, one gets 
\begin{equation}
    D^B D_A D_B m = D_A D^2 m + \tensor{R}{_A^C}D_C D^2 m = D_A (D^2 m + 1) m \ ,
\end{equation}
so it is obvious that the first two terms in \nr{div1} give a total gradient contribution. Now suppose that the bracket term in \nr{div1} also is a total gradient:
\begin{equation}
    D_A \psi = \frac{1}{2}\left(\eps_{AC} D^B D^C D_B \bar{m} + \eps_{BC} D^B D^C D_A \bar{m}\right) \ ,
\end{equation}
for some scalar function $\psi$. This implies that
\begin{equation}
    D^2 \psi = \frac{1}{2}\left(\eps_{AC} D^A D^B D^C D_B \bar{m} + \eps_{BC} D^A D^B D^C D_A \bar{m}\right) = 0 \ ,
\end{equation}
where the RHS's vanishing was shown above. This means that $\psi$ must be a constant, which in turn implies that
\begin{equation}
    \eps_{AC} D^B D^C D_B \bar{m} + \eps_{BC} D^B D^C D_A \bar{m} = 0 \ .
\end{equation}
Commuting covariant derivatives in both terms yields
\begin{equation}
    \eps_{AC} D^C D^2 \bar{m} + \eps_{AC} R^{CE} D_E \bar{m} + \frac{1}{2}\eps^{BC}\tensor{R}{_B_C_A^E} D_E \bar{m} = 0 \ .
\end{equation}
Again, we use the fact that the Ricci tensor equals the metric to simplify the second term, and one can also show that
\begin{equation}
    \eps^{BC}\tensor{R}{_B_C_A^E} = 2\tensor{\eps}{_A^E} \ ,
\end{equation}
so we get that
\begin{equation}
    D_A (D^2 + 2) \bar{m} = 0 \ .
\end{equation}
This of course implies that $(D^2 + 2) \bar{m}$ is a constant. Expanding it in spherical harmonics yields
\begin{equation}
    \sum_{l,m} \bar{m}_{lm} (2 - l(l+1)) Y_{lm} = \text{const.} \equiv C_{00} Y_{00}
\end{equation}
which implies that 
\begin{equation}
    \bar{m}_{lm} = 0 \quad \forall l > 1 \ .
\end{equation}
Hence, only the $l=0,1$ modes of $\bar{m}$ could potentially contribute to $m\ABD$. However, by a direct computation one can show that
\begin{equation}
    \eps_{AC} D^C D_B Y_{lm} + \eps_{BC} D^C D_A Y_{lm} = 0
\end{equation}
for $l=0,1$ and $m = -l,\ldots,l$. Therefore, the co-closed part completely vanishes, when we demand that $m\ABD$ satisfies \nr{div1}, and the memory tensor becomes
\begin{equation}
    m\ABD = D_A D_B m - \frac{1}{2}h\ABD D^2 m \ .
\end{equation}
Above, we learned that the $l=1$ mode of $\phi$ vanishes. From \nr{final1} and \nr{D2phi}, on the other hand, we further find that
\begin{equation}
    \frac{1}{2}D^2(D^2 + 2) m = \Delta P - 8 \pi F \Omega^{-2} \left( \Omega + r \pt_r \Omega + r \pt_\eta \Omega \right)
\end{equation}
Spherical harmonic expansion yields
\begin{equation}
    -\frac{1}{2}\sum_{l,m} l(l+1)(2-l(l+1)) m_{lm} Y_{lm} = \Delta P - 8 \pi F \Omega^{-2} \left( \Omega + r \pt_r \Omega + r \pt_\eta \Omega \right)
\end{equation}
which can be integrated mode by mode to find the coefficients $m_{lm}$. Therefore, the solution for the memory tensor reads
\begin{equation}
    m\ABD = \sum_{l\geq 2} m_{lm} \left(D_A D_B Y_{lm} - \frac{1}{2}h\ABD D^2 Y_{lm} \right) \ ,
\end{equation}
with
\begin{equation}
    m_{lm} = \frac{2}{(l-1)l(l+1)(l+2)}\int d\Omega \, Y^*_{lm} (\Delta P - 8\pi F \Omega^{-2} \left( \Omega + r \pt_r \Omega + r \pt_\eta \Omega \right)) \ .
\end{equation}
Notice that here the sum over modes starts from $l=2$; the above calculation shows that $l=0,1$ modes do not contribute to the memory tensor.

\subsection{Vanishing of further contributions to the decomposition of memory tensor}\label{app:proof}
We will review the argument \cite{Higuchi:1986wu} that the symmetric, traceless, and divergence-free tensor vanishes on ${\mathbb{S}}^2$ which would otherwise appear in the decomposition of the memory tensor (\ref{hodgedec}).

Let $S\ABD$ be a symmetric, traceless, and divergence-free tensor on ${\mathbb{S}}^2$. Then we can write
\begin{equation}
    D_{[A}S_{B]C} = \frac{1}{2}\eps\ABD \eps^{EF}D_E S_{FC} \equiv \eps\ABD V_C \ .
\end{equation}
Contracting with $h^{BC}$, the LHS vanishes due to $S\ABD$ being traceless and divergence-free, which implies that $V_C = 0$ and this in turn entails that
\begin{equation}
    D_{[A}S_{B]C} = 0 \ .
\end{equation}
Now, using this and the Ricci identity, we obtain
\begin{equation}
    D^C D_C S\ABD = D^C D_A S_{CB} = D_A D^C S_{CB} + \tensor{R}{^C_A_C^E}S_{EB} + \tensor{R}{^C_A_B^E}S_{CE} \ .
\end{equation}
The divergence term on the RHS vanishes, and using the simple form \nr{maxsym} for the Riemann tensor,
we get
\begin{equation}
    D^C D_C S\ABD = 2 S\ABD \ .
\end{equation}
Then, contracting both sides with $S\ABU$ and rearranging the derivatives on the LHS, we have
\begin{equation}
    D^2 f - 2 D_C S\ABD D^C S\ABU = 4 f \ , \quad f\equiv S\ABD S\ABU \ .
\end{equation}
The second term on the LHS is non-positive so
\begin{equation}
    D^2 f \geq 4 f \geq 0 \ .
\end{equation}
But ${\mathbb{S}}^2$ is a compact manifold so the image $f({\mathbb{S}}^2)$ of the map $p \mapsto f(p)$ is a compact subset of $\mathbb{R}$; thus, $f$ must have a maximum in its image. Then the Hopf maximum principle implies that, since $D^2 f \geq 0$, $f$ has to be a constant. But then $0 = D^2 f \geq f \geq 0$, which means that $S\ABD S\ABU = f = 0$. This can only be the case if $S\ABD = 0$.

\bibliographystyle{JHEP}
\bibliography{refs}

\end{document}